\documentclass[prl,aps,showpacs,twocolumn,unsortedaddress]{revtex4-1}

\usepackage{graphics, bm, xspace}
\usepackage{graphicx}
\usepackage{psfrag}
\usepackage{amsmath}
\usepackage{amssymb}
\usepackage{epsfig}
\usepackage[margin=0pt]{caption}
\usepackage{subcaption}
\usepackage{grffile}
\usepackage{times}
\usepackage{color}
\usepackage{multirow}
\usepackage[bookmarks=false,linkcolor=blue,urlcolor=blue,colorlinks,citecolor=blue]{hyperref}
\usepackage{float}

\graphicspath{{./}}

\newcommand{\eps}    {\epsilon}

\newcommand{\bb}     {{\bf b}}

\newcommand{\kk}     {{\bf k}}

\newcommand{\qq}     {{\bf q}}

\newcommand{\KK}     {{\bf K}}

\newcommand{\GG}     {{\bf G}}

\newcommand{\stau}     {{\boldsymbol\tau}}

\newcommand{\degree}{^\circ}

\newcommand{\stkout}[1]{\ifmmode\text{\sout{\ensuremath{#1}}}\else\sout{#1}\fi}

\newcommand{\Ds}{D^{s}_{xx}}
\newcommand{\dsc}{d^{s,\rm conv}_{xx}}
\newcommand{\Dsc}{D^{s,\rm conv}_{xx}}
\newcommand{\dsg}{d^{s,\rm geom}_{xx}}
\newcommand{\Dsg}{D^{s,\rm geom}_{xx}}

\begin{document}

\title{Geometric and conventional contribution to superfluid weight in twisted bilayer graphene}
\author{Xiang Hu$^1$, Timo Hyart$^2$,  Dmitry I. Pikulin$^3$, Enrico Rossi$^1$}
\affiliation{
             $^1$Department of Physics, William \& Mary, Williamsburg, VA 23187, USA,\\
             $^2$International Research Centre MagTop, Institute of Physics, Polish Academy of Sciences, Aleja Lotnikow 32/46, PL-02668 Warsaw, Poland, \\
             $^3$Microsoft Quantum, Microsoft Station Q, University of California, Santa Barbara, California 93106-6105
            }
\date{\today}
   
\begin{abstract}
By tuning the
angle between graphene layers to specific ``magic angles'' the lowest energy bands of twisted bilayer graphene (TBLG) can be made
flat. The flat nature of the bands favors the formation of collective ground states and, in particular,
TBLG has been shown to support superconductivity. 
When the energy bands participating in the superconductivity are well-isolated, 
the superfluid weight
scales inversely with the effective mass of such bands.
For flat-band systems one 
would therefore conclude that even if superconducting pairing is present most of the signatures of the superconducting state 
should be absent. This conclusion
is at odds with the experimental observations for TBLG.
We calculate the superfluid weight for TBLG taking into account both the conventional contribution 
and the 
contribution arising from the 
quantum geometry of the bands. 
We find that both contributions are larger than one would expect treating the bands as well-isolated,
that at the magic angle the geometric contribution is larger than the conventional 
one,
and that 
for small deviations away from the magic angle the conventional 
contribution is larger than the geometric one.
Our results show that, despite the flatness of the bands the superfluid weight in TBLG is finite and consistent with 
experimental observations. We also show how the superfluid weight can be tuned by varying the chemical potential and the twist angle
opening the possibility to tune the nature of the superconducting transition between the standard BCS transition
and the Berezinskii-Kosterlitz-Thouless transition.
\end{abstract}

\maketitle

The ability to control accurately the twist angle $\theta$ between two-dimensional crystals forming
a van der Waals
systems~\cite{Li2010, Geim2013,Neto2016,Kim2016} has
recently emerged as a powerful way to tune the electronic
properties of a condensed matter system. The most remarkable example of such tunability has been observed in twisted bilayer graphene (TBLG). 
For most values of the twist angle between the graphene sheets the systems behave as a normal two-dimensional (2D) semimetal,
however for specific ``magic angles''~\cite{dossantos2007,Mele2010,Morell2010,Bistritzer2011a,Magaud2012} 
the system's lowest energy bands become almost completely flat and the system may support topological properties \cite{Song2019,Bernevig2018,Po2019,Dai2019,Yang2019}. 
Quenched kinetic energy in the flat bands increases the importance of interactions and leads to superconductivity
and other correlated states~\cite{Kopnin2011, Ojajarvi2018, Wu2018, PeltonenPhysRevB.98.220504, GuoPhysRevB.97.235453,LiuPhysRevLett.121.217001,
Ashvin2019, Senthil2018, TangPhysRevB.99.094521, RoyPhysRevB.99.121407, Fu2019, WuPhysRevB.99.195114, Wu2019} recently observed in graphene moir\'e superlattices ~\cite{Kim2017,Cao2018,Cao2018b,Yankowitz2019, Chen2019, Lu2019arXiv,Shen2019arXiv190306952S,Kim2019,  Cao2019arXiv190308596C, Chen2019arXiv190506535C}.
The hallmark signature of the superconducting state is the absence of electrical resistance.
For this to happen the superfluid weight
$\Ds$ must be nonzero. 
For an isolated parabolic band at zero temperature  $\Ds \propto n/m^*$, where $n$ is the electron density,
and $m^*$ the effective electron's mass. From this expression one would conclude that the standard signature of superconductivity might be absent
for flat bands because one expect $1/m^*$ to vanish proportionally to the bandwidth.
This is not what happens experimentally in TBLG.

In order to reconcile experimental observations and theory we notice
that the above expression for $\Ds$ assumes an isolated band and neglects the interband matrix elements of the current operator. Neither of these assumptions is valid in TBLG. In an isolated band the density of electrons within the band is constant, and therefore when the superconducting transition occurs the chemical potential is renormalized. The superfluid weight depends strongly on the chemical potential and this renormalization is responsible for the appearance $1/m^*$ dependence of the intraband (conventional) contribution to the superfluid weight. In a semimetal where both electron- and hole-like bands are present, such as TBLG, the densities in each band are not conserved in the transition separately and the dependence on the chemical potential is weak so that the conventional contribution can be much larger than expected for isolated bands. Moreover, the bandwidth of
the low-energy bands, even though very small, is still finite and larger than
the superconducting gap. Therefore, the velocity can be large at some points of the Brillouin zone further enhancing the conventional contribution to the weight. On the other hand, the interband matrix elements give rise to the so called geometric part of $\Ds$, which can be large even for completely flat band well isolated from other bands~\cite{Peotta2015,Liang2017}.

In this work we calculate the superfluid weight of superconducting twisted bilayer graphene (TBLG)
taking into account both the conventional and the geometric parts. 
We assume singlet pairing
and use the experimentally measured value of $T_c$ to set the value of the
coupling constant that enters the mean field gap equation.
We obtain the dependence of the superconducting weight on the twist angle and separate
the conventional and geometric parts. 
We find that at one of the ``magic angles'', $\theta =1.05\degree$, 
the geometric contribution is approximately twice as large
as the conventional one. However, just off the magic angle the conventional contribution is larger than the geometric one. We also obtain the dependence of the Berezinski-Kosterlitz-Thouless $T_{\mathrm{KT}}$ temperature
on $\theta$ and show that its scaling with the chemical potential is different at the magic angle and away from it. Because our calculations take into account the full band structure of TBLG and include both intra- and interband contributions, they can be used for quantitative predictions and they go beyond the models and approximations previously used in deriving bounds for the superfluid weight \cite{Hazra2018, Fang2019}.

To model the TBLG we use the approach described in Ref.~\onlinecite{Bistritzer2011a,Wu2018}. The low-energy states of the isolated single layers of graphene
 are located at the $\KK$ and $\KK'= -\KK$ valleys of the Brillouin zone (BZ). Close to $\KK$ the Hamiltonian for each layer $l=\pm 1$  is
\begin{equation}
H_{\KK, l}(\kk)= e^{-il\frac{\theta}{4}\tau_z}[\hbar v_F(\kk-{\boldsymbol\kappa}_l)\cdot\stau -\mu\tau_0]e^{il\frac{\theta}{4}\tau_z},
\end{equation}
where  $v_F=10^6$~m/s is graphene's Fermi velocity, $\mu$ is the chemical potential,
and ${\tau_i}$ ($i=0,1,2,3$) are the $2\times2$ Pauli matrices in sublattice space. Because of the rotation of each layer by angle $\theta/2$ the Dirac cone position in layer $l$ is shifted to ${\boldsymbol\kappa}_l$. We choose moir\'e BZ in which ${\boldsymbol\kappa}_l$ are located at the corners and refer to the center of this BZ as the $\gamma$ point.
This leads to a Hamiltonian for TBLG around the $\KK$ point
\begin{equation}
    H_{TBL,\KK}=\begin{pmatrix} H_{\KK, +1} & T(\mathbf{r})\\ T^\dag(\mathbf{r}) & H_{\KK, -1} \end{pmatrix},
\end{equation}
with periodically varying interlayer tunneling terms 
$T({\bf r})=w[T_0+e^{-i{\bf b}_2\cdot{\bf r}}T_{+1}+e^{-i({\bf b}_2-{\bf b}_1)\cdot{\bf r}}T_{-1}]$, where $T_j=\tau_0+\cos(2\pi j/3)\tau_x+\sin(2\pi j/3)\tau_y$, $\bb_1=(\sqrt{3}Q,0)$ and
$\bb_2 =(\sqrt{3}Q/2,3Q/2)$ are reciprocal basis vectors, $Q=\frac{8\pi}{3 a_0}\sin(\theta/2)$, $a_0$ is the lattice constant of graphene and $w=118$ meV \cite{Jung2014,Wu2018}. $H_{\KK'}$ is obtained from $H_{\KK}$ via time-reversal.

\captionsetup{justification=centerlast, indention=0cm, singlelinecheck=true}
\begin{figure}[htb]
  \centering 
  \includegraphics[width=\columnwidth]{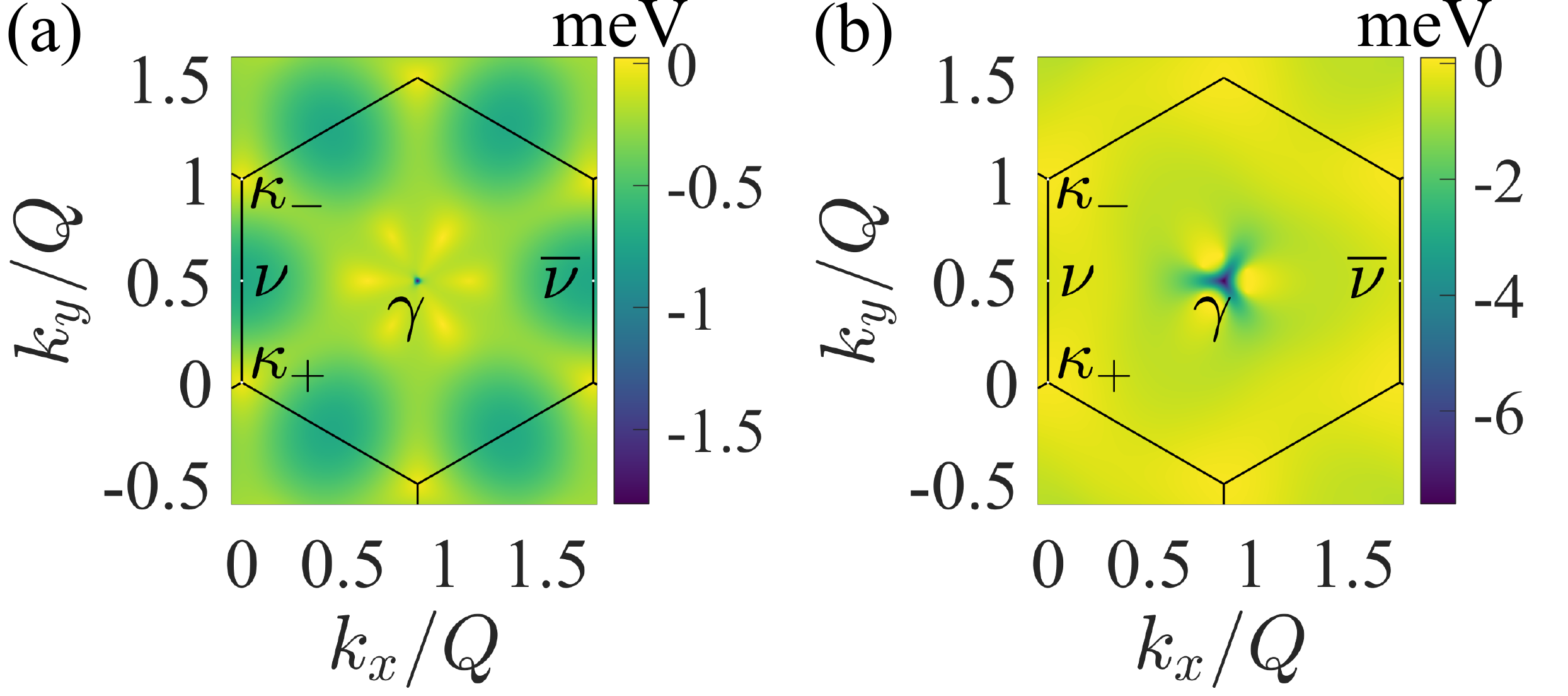}
  \caption{The dispersion of the lower flat band of TBLG for (a) $\theta=1.05\degree$ and (b) $\theta=1.00\degree$. The high symmetric points in the moir$\acute{\text{e}}$ Brillouin zone (BZ) are also shown.
\hfill  \label{fig:bands}}
\end{figure} 

We leave $d$-wave pairing\cite{Wu2018} for the future studies and focus on $s$-wave pairing. In the presence of superconductivity the mean field theory in Nambu space is described by the
Bogoliubov-de-Gennes Hamiltonian
\begin{eqnarray}
 H_{BdG}  = \left[
 \begin{array}{cc}
   H_{TBL, \KK}(\kk)          &  \hat\Delta_s \\
  \hat\Delta_s^\dag            &  -H^T_{TBL, \KK'}(-\kk)                      
\end{array}
\right],
\label{eq:HsKSC}
\end{eqnarray}
and 
$\hat{\Delta}_s=\Delta\tau_0\sum_{\bb}\Delta_\bb e^{i{\bf b}\cdot{\bf r}}$, where 
$\Delta$ is the overall amplitude of the superconducting gap, and $\Delta_\bb$ is the normalized coefficient of the
$\bb=m_1\bb_1 + m_2\bb_2$ ($m_i\in Z$) Fourier component. 
In the remainder we assume $\Delta=1.764 k_B T_c$, and determine $T_c$ and the coefficients $\Delta_\bb$
by solving the linearized gap equation\cite{Wu2018,SM}.

Using standard linear response theory we can obtain the expression for the superconducting weight \cite{Peotta2015, Liang2017,SM}
\begin{eqnarray}\label{eq:StiffTimo}
D^s_{\mu\nu}&=&\sum_{{\bf k},i,j}\frac{n(E_j)-n(E_i)}{E_i-E_j}
\left(\frac{1}{4L^2}\langle\psi_i|\hat{v}_{\mu}|\psi_j\rangle
\langle\psi_j|\hat{v}_{\nu}|\psi_i\rangle \right.
\nonumber\\
&&-\left.\frac{1}{L^2}\langle\psi_i|\hat{v}_{cf,\mu}|\psi_j\rangle
\langle\psi_j|\hat{v}_{cf,\nu}|\psi_i\rangle\right),
\end{eqnarray}
where $L\times L$ is the size of the two dimensional system, $n(E)$
is the Fermi distribution function, $E_{i}$, $|\psi_i(\kk)\rangle$ are the  eigenvalues and eigenvectors of $H_{BdG}$,
and $\mu,\nu=x,y$ represent the directions. In the remainder, we focus on the case $\nu=\mu$. 
We have the velocity operators
$\hat{v}_{\mu}({\bf k})=\partial H_{BdG}/\partial k_{\mu}$,
$\hat{v}_{cf,\mu}({\bf k})=(1/2)\gamma_z\partial H_{BdG}/\partial k_{\mu}$, 
($\gamma_z$ is the Pauli matrix acting in Nambu space).

Let $H_+$ and $H_-$ be the particle and hole Hamiltonians, respectively, of $H_{BdG}$,
$|\psi_{\pm m}\rangle$ the eigenstates of $H_\pm$,
$w_{\pm im}\equiv \langle\psi_{\pm m}|\psi_i\rangle$,
and
$v^+_{\mu}\equiv \partial_{\mu}{H_+}$,
$v^-_{\mu}\equiv -\partial_{\mu}{H_-}$.
In terms of these quantities we have~\cite{Liang2017}:
%
%
\begin{eqnarray}\label{eq:proj}
D^s_{\mu\mu}&=&\frac{1}{L^2}\sum_{{\bf k}ijmnpq}\frac{n(E_i)-n(E_j)}{E_i-E_j}\nonumber\\
&&w^*_{+im}v^+_{\mu mn}w_{+jn}
w^*_{-jp}v^-_{\mu pq}w_{-iq},
\end{eqnarray}
where $m,n$ and $p,q$ index the particle and hole bands. The matrix elements with $m\neq n$ and $p\neq q$ in Eq.~(\ref{eq:proj}) represent pure interband contribution. By defining
\begin{eqnarray*}
V^d_{\pm\mu ij}&\equiv&\sum_{m}w^*_{\pm im}v^{\pm}_{\mu mm}w_{\pm jm},\nonumber\\
V^o_{\pm\mu ij}&\equiv&\sum_{m\neq n}w^*_{\pm im}v^{\pm}_{\mu mn}w_{\pm jn},
\end{eqnarray*}
we can separate Eq.~(\ref{eq:proj}) into a conventional and a geometric part: 
\begin{eqnarray}
D^{s,\mathrm{conv}}_{\mu\mu}&=&\frac{1}{L^2}\sum_{{\bf k}ij}\frac{n(E_i)-n(E_j)}{E_i-E_j}\nonumber\\
&&(V^d_{+\mu ij}V^d_{-\mu ji}+V^d_{+\mu ij}V^o_{-\mu ji}+V^o_{+\mu ij}V^d_{-\mu ji}) \nonumber\\
D^{s,\mathrm{geom}}_{\mu\mu}&=&\frac{1}{L^2}\sum_{{\bf k}ij}\frac{n(E_i)-n(E_j)}{E_i-E_j}
V^o_{+\mu ij}V^o_{-\mu ji}.\label{eq:Stiffness_Integral}
\end{eqnarray}
Below we show that both the conventional~\cite{Scalapino1992,Scalapino1993} and the
geometric contribution~\cite{Peotta2015,Liang2017}
are important for the superfluid weight in TBLG.

\captionsetup{justification=centerlast, indention=0cm, singlelinecheck=true}
\begin{figure}[htb]
 \begin{center}
  \centering 
  \includegraphics[width=\columnwidth]{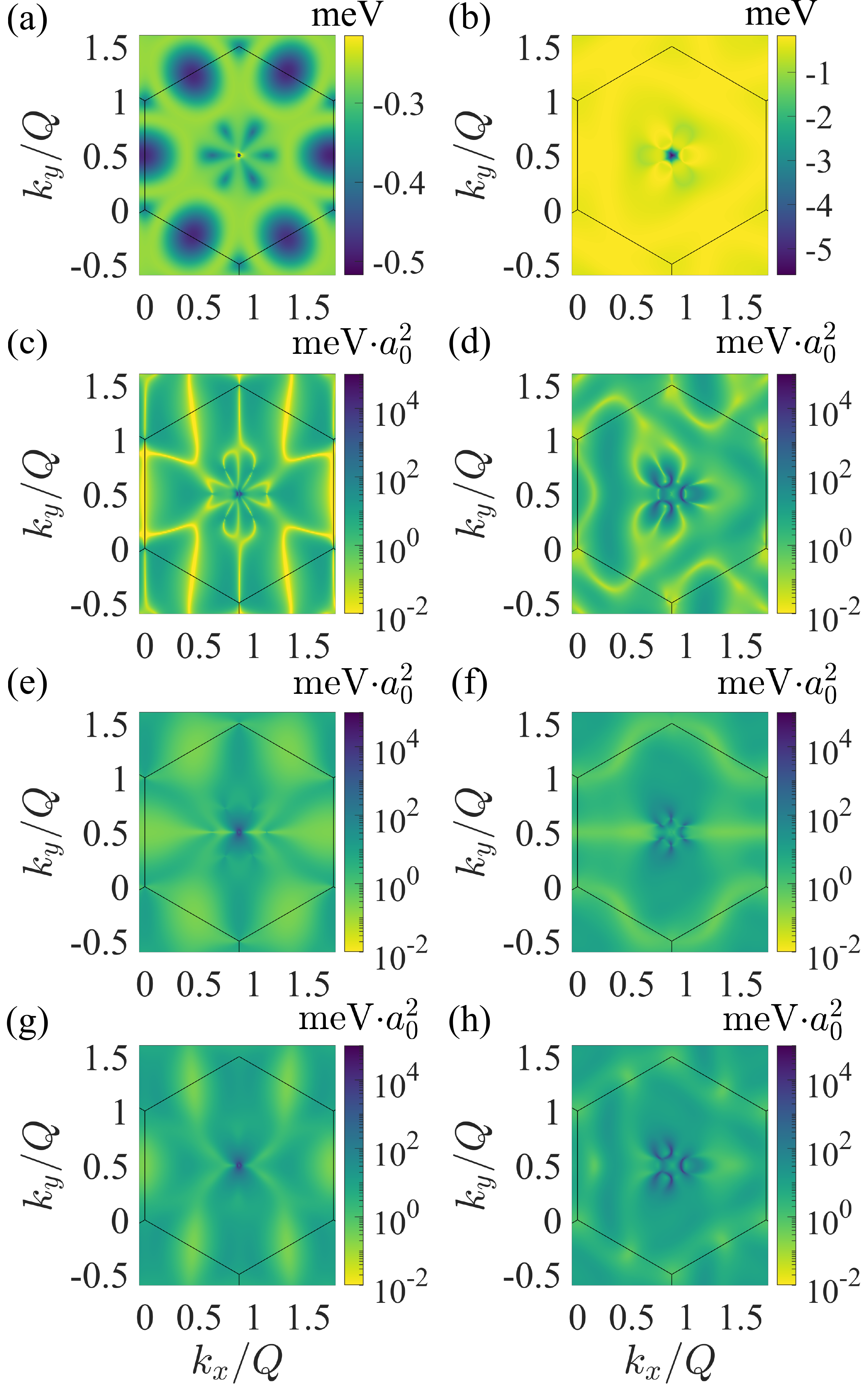}
  \caption{
          The dispersion of superconducting band and superfluid weight integrand. 
          Left column $\theta=1.05\degree$, and right column $\theta=1.00\degree$.
          (a,b) the lowest quasiparticle bands with superconducting gap.
          (c,d) $\dsc(\kk)$.
          (e,f) $\dsg(\kk)$.
          (g,h) $d^{s,\rm{total}}_{xx}(\kk)$. All the figures are obtained with $\mu=-0.30$ meV and $T_c=1.6305$~K $T_c=1.2119$~K for
          $\theta=1.05\degree$, $\theta=1.00\degree$, respectively.    
        \hfill 
  \label{fig:s-wave}}
 \end{center}
\end{figure}

Figure~\ref{fig:bands} shows the dispersion of the lower flat band without superconductivity. It is plotted 
in the moir\'e BZ 
for two different values of $\theta$: $\theta=1.05\degree$, the magic angle, and $\theta=1.00\degree$.  
For each angle we see 
a sharp feature in the dispersion at the $\gamma$ point, which is away from zero energy. 
For $\theta=1.05\degree$
the bandwidth of the nearly flat moir\'e band is about 2~meV, whereas for $\theta=1.00\degree$ it is around 5~meV. 
The bandwidth of the lowest energy bands and the value of the magic angle can differ between experiments~\cite{Kerelsky2019}, a fact
that can be taken into account by tuning $w$, see \cite{SM}.
For $\theta=1.00\degree$  we see that the bands exhibit deep and narrow valleys,  green regions emanating from the $\gamma$ point.
Around these valleys the quasiparticle energy $\eps(\kk)$ varies rapidly with $\kk$ producing high local velocity despite the fact that the bandwidth is only few meVs.

Figures~\ref{fig:s-wave}~(a),~(b) show the profile of $\eps(\kk)$ for the lowest excitation
in the presence of s-wave pairing, for $\theta=1.05\degree$, and $\theta=1.00\degree$.
The amplitude and Fourier components of the superconducting gap are obtained by solving the mean field gap equation~\cite{SM}.
We see that also in the presence of a superconducting gap the bands exhibit the same qualitative
features as the bands with no pairing~Fig~\ref{fig:bands}.

Figures~\ref{fig:s-wave}~(c),~(d) show the momentum space profile of the integrand, $d^{s,\rm conv}_{xx}(\kk)$, that enters the expression \eqref{eq:Stiffness_Integral}
for $\Dsc$ for $\theta=1.05\degree, 1.00\degree$, respectively.
We see that for $\theta=1.05\degree$ $\dsc$  is peaked at $\gamma$ point, and is otherwise quite uniform and small. 
At bands crossings $d_{xx}^{s,{\rm geom}}$ is expected to be large as long as the Berry curvature is not zero,
regardless of the nature of the crossing~\cite{Rhim2019}.
For $\theta=1.00\degree$ $\dsc$ is strongly peaked at the position of the valleys that we identified
in Fig.~\ref{fig:bands}~(b). This clearly shows that the conventional contribution to $D^s$ can depend
very strongly on the twist angle and in general cannot be assumed to be negligible despite the smallness of the bandwidth.
The reason is that even for narrow bands, the expectation value of the velocity operators can be 
non-negligible.
Figure.~\ref{fig:s-wave}~(e),~(f) show the profile of  the integrand, $\dsg(\kk)$, that enters the expression
of $\Dsg$ for the same conditions used to obtain panels (c)~and~(d).
For $\theta=1.05\degree$ $\dsg(\kk)$ is strongly peak at the $\gamma$ point and 
on average is larger
than the conventional term. 
This shows that at the magic angle the geometric contribution to $D^s_{xx}$ is significant and larger than the conventional contribution.
For $\theta=1.00\degree$,
however, $\dsc(\kk)$ is large in most of the moir\'e BZ
so that the conventional contribution to $\Ds$ is larger than the geometric one.
As the bands become flatter the conventional contribution, for fixed electron's density, decreases and so we can expect its importance to
decrease relative to the geometric contribution.
Fig.~\ref{fig:s-wave}~(g),~(h) show the sum $\dsc(\kk)+\dsg(\kk)$.
%
%
It is worth pointing out that the spin Chern number, $C$, of the lowest energy bands is zero, but in general $\Dsg$
is nonzero even when $C=0$~\cite{Liang2017}.
%
%

\begin{figure}[htb]
 \begin{center}
  \centering 
  \includegraphics[width=\columnwidth]{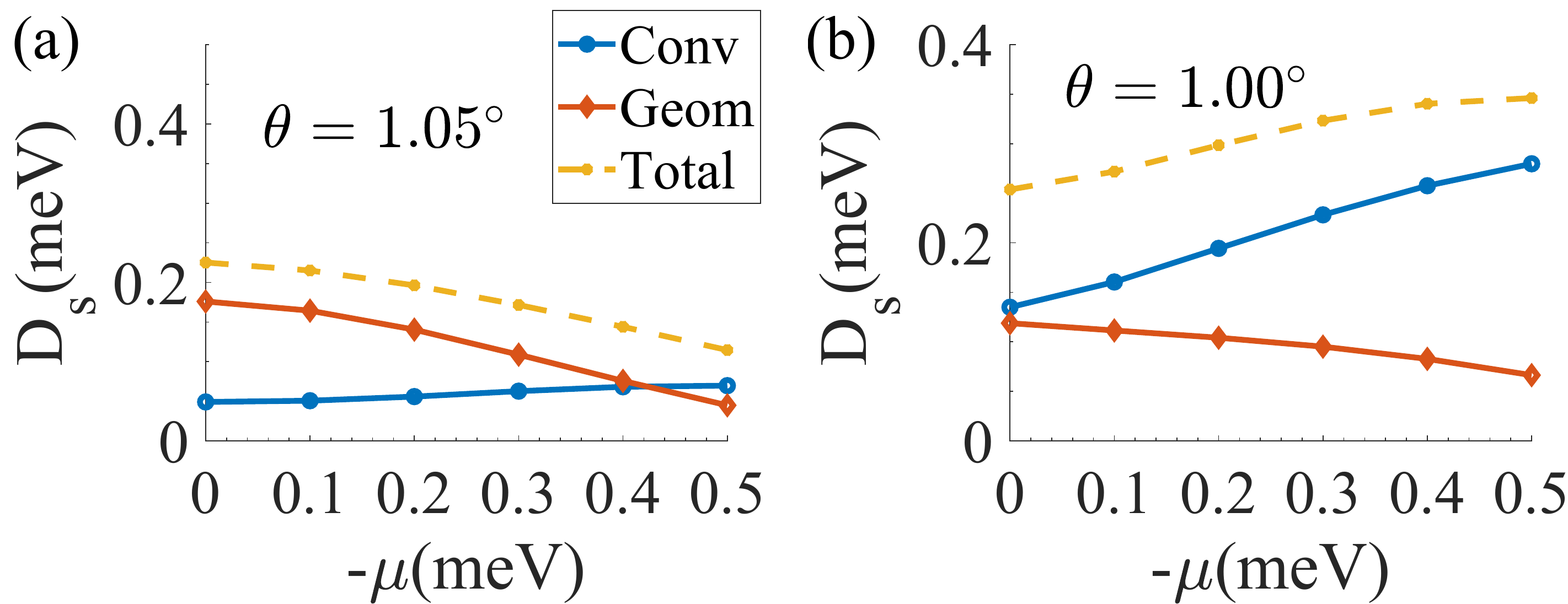}
  \caption{
          $D_s$ as a function of the chemical potential for (a) $\theta=1.05\degree$ and (b) $\theta=1.00\degree$ at $\Delta=\Delta(\mu)$.
  \hfill
  \label{fig:rhos_vs_mu}}
 \end{center}
\end{figure}

We continue by obtaining the dependence of $\Dsc$, and $\Dsg$ on the chemical potential.
From the initial discussion we expect $\Dsc$ to increase with the electron density and therefore with $|\mu|$.
The scaling of $\Dsg$ with respect to $\mu$ depends on the details of the quantum metric of the bands~\cite{Liang2017}.
Fig.~\ref{fig:rhos_vs_mu} shows the evolution of $\Dsc$, $\Dsg$, and $\Ds$ with $\mu$
for the cases of $\theta=1.05\degree$ and $\theta=1.00\degree$.
To obtain these results the superconducting gap is obtained for each value of $\mu$.
The results of Fig.~\ref{fig:rhos_vs_mu} confirm the expectation that $\Dsc$ increases with $|\mu|$, for both the magic angle and 
$\theta=1.00\degree$. They also show that for both angles the geometric contribution decreases with $|\mu|$.
Considering that $\Ds$ controls the critical temperature, $T_{\mathrm{KT}}$, for the Berezinskii-Kosterlitz-Thouless (BKT)
phase transition~\cite{Berezinski1971,Kosterlitz1973}, the results of Fig.~\ref{fig:rhos_vs_mu} show that in TBLG
it could be possible in principle to tune the nature of the transition, BCS, or BKT by simply tuning the chemical potential.
\captionsetup{justification=centerlast, indention=0cm, singlelinecheck=true}

\begin{figure}[htb]
 \begin{center}
  \centering 
  \includegraphics[width=\columnwidth]{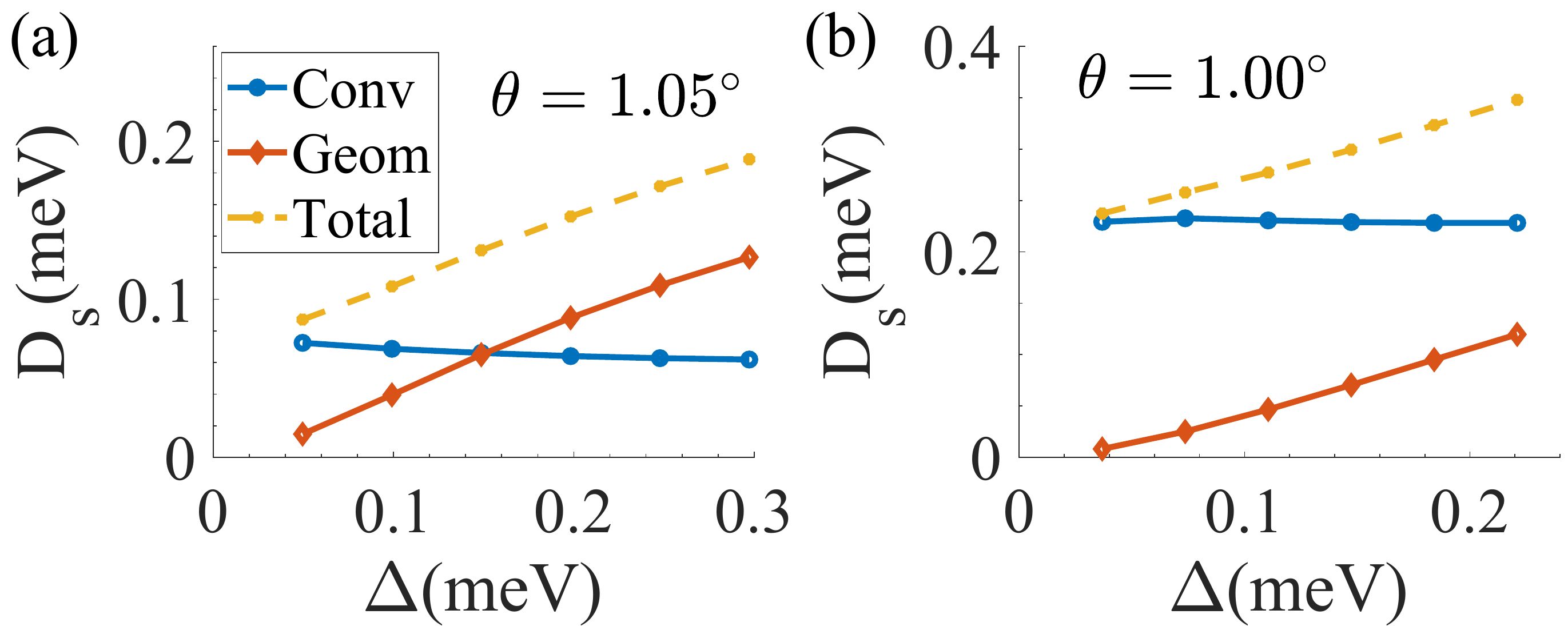}
  \caption{
          $D_s$ as a function of $\Delta$ for $\theta=1.05\degree$, (a), and $\theta=1.00\degree$ (b).
          $\mu=-0.3$~meV. \hfill
  \label{fig:rhos_vs_Delta}}
 \end{center}
\end{figure}

An increase of $\Delta$, keeping $\mu$ fixed, is expected to cause an increase of $D^s_{xx}$.
This is confirmed by the results of Fig.~\ref{fig:rhos_vs_Delta}. Again, we can see at $\theta=1.05\degree$ the geometric contribution 
is significant while at $\theta=1.00\degree$ the conventional contribution dominates.

\captionsetup{justification=centerlast, indention=0cm, singlelinecheck=true}

\begin{figure}
 \begin{center}
  \centering 
  \includegraphics[width=\columnwidth]{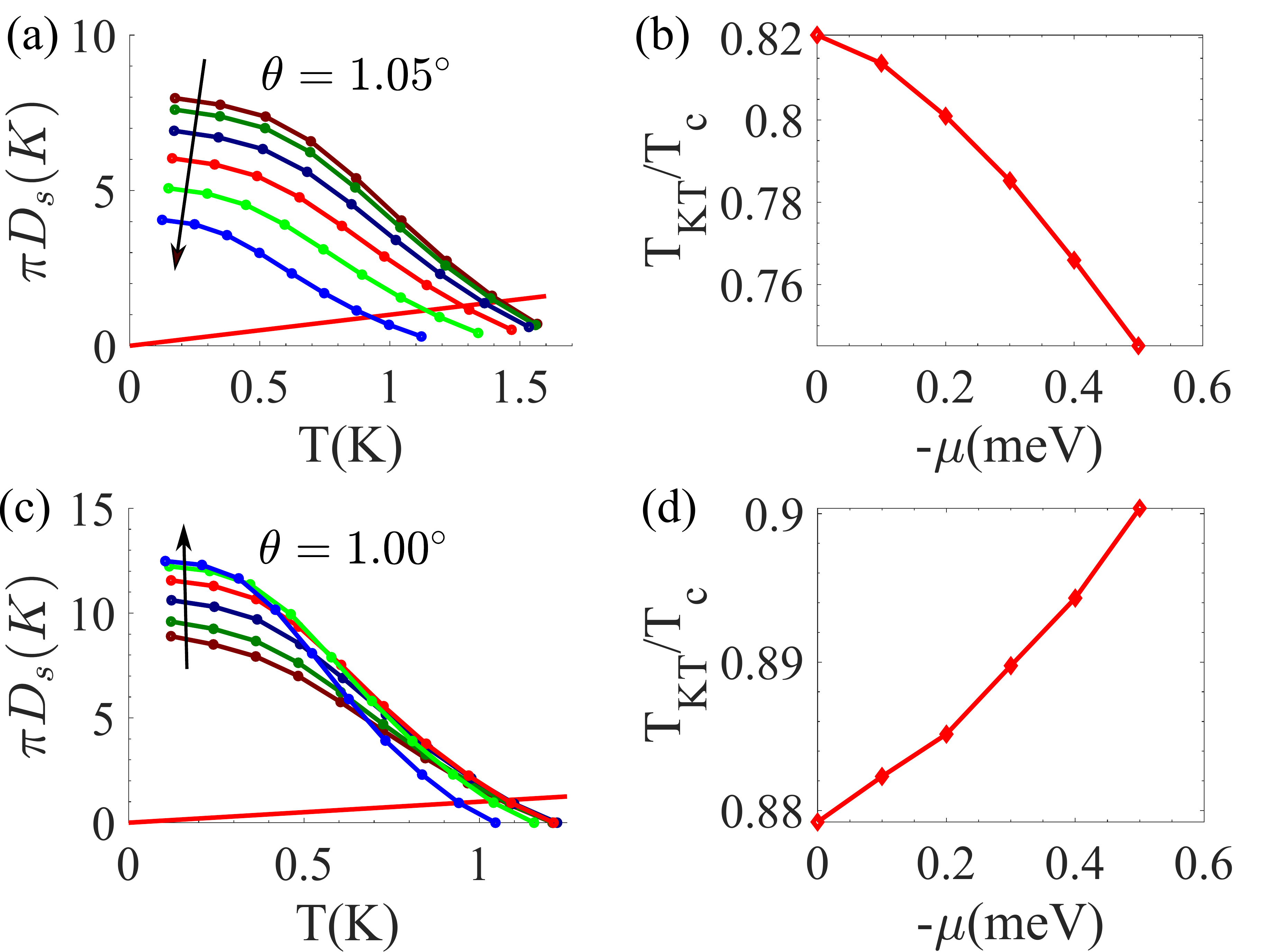}
  \caption{
          (a) $D^s$ versus $T$ for $\theta=1.05\degree$ and different values of $\mu$: $\mu$ goes from 0~meV to -0.5~meV along the direction of the arrow.
          (b) $T_{\mathrm{KT}}/T_c$ as function of $\mu$ for $\theta=1.05\degree$.
          (c) Same as (a) for $\theta=1.00\degree$.
          (d) $T_{\mathrm{KT}}/T_c$ as function of $\mu$ for $\theta=1.00\degree$. 
        \hfill
  \label{fig:TKT_vs_mu}}
 \end{center}
\end{figure}

In Figs.~\ref{fig:TKT_vs_mu}~(b),~(d) we show the BKT transition temperature as a function of $\mu$ obtained from the equation 
$k_B T_{\mathrm{KT}}=\pi D^s(\Delta(T_{\mathrm{KT}}),T_{\mathrm{KT}})$, assuming $\Delta(T)=1.764k_B T_c\sqrt{1-T/T_c}$. 
The pre-factor on the r.h.s. of the equation for $T_{\mathrm{KT}}$ is twice $\pi/2$ due to the valley degeneracy.
In Figs.~\ref{fig:TKT_vs_mu}~(a),~(c) the curves with solid circles show the evolution of $D^s(\Delta(T),T)$ with $T$
for different values of $\mu$ for  $\theta=1.05\degree$ and $\theta=1.00\degree$, respectively.
The intersection of these curves with the solid line $k_BT$ returns the values of $T_{\mathrm{KT}}(\mu)$.
We note that $T_{\mathrm{KT}}$ is fairly close to $T_c$, well above the lower bound set by previous studies~\cite{Fang2019}.

\begin{figure}[htb]
 \begin{center}
  \centering 
  \includegraphics[width=\columnwidth]{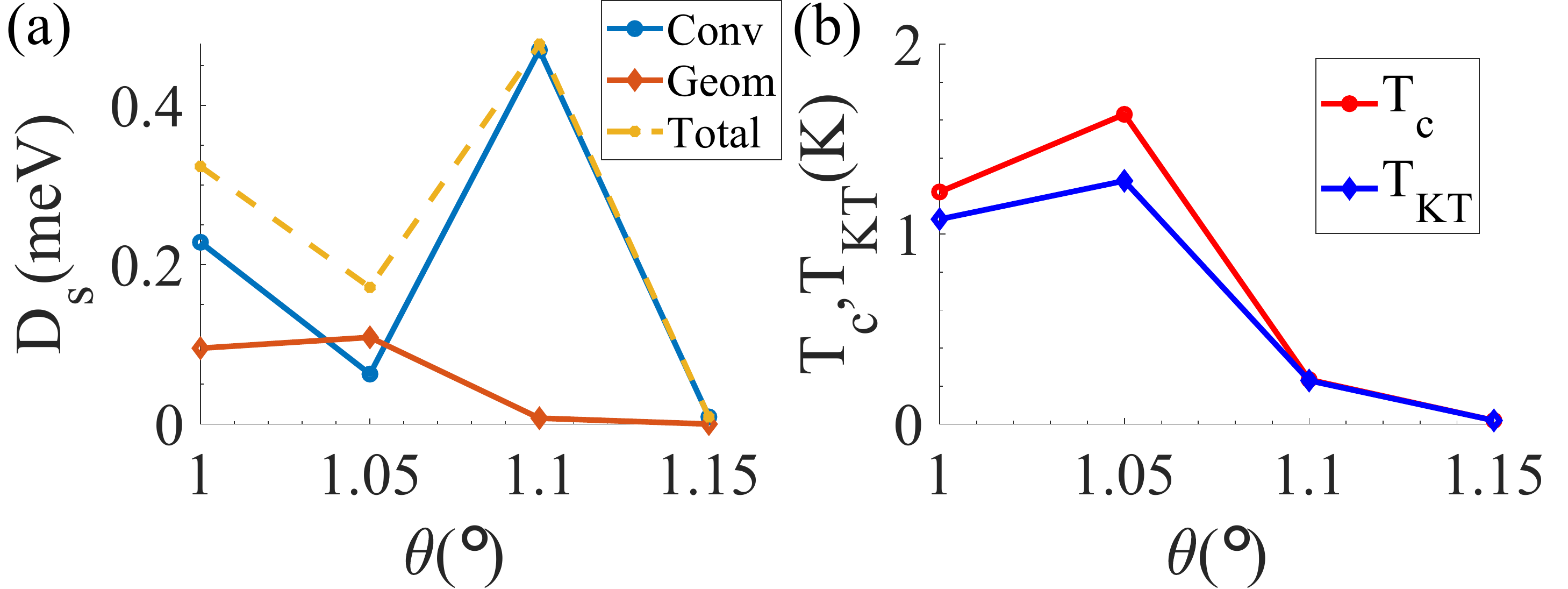}
  \caption{
          (a)$D_s$, and (b) $T_c$ and $T_{\mathrm{KT}}$ as functions of twist angle. $\mu=-0.3$~meV.
         \hfill 
  \label{fig:Ds_vs_theta}}
 \end{center}
\end{figure}

We obtained the value of $\Ds$ ($\Dsc$, $\Dsg$) for different twist angles using the
corresponding values of $T_c$. The results are shown in 
Fig.~\ref{fig:Ds_vs_theta}~(a). We see that despite the fact that $T_c$ is lower for $\theta=1.10\degree$
than for $\theta=1.05\degree$ the superconducting weight is larger for $\theta=1.10\degree$.
This is because for $\theta=1.10\degree$ the conventional contribution to $\Ds$ is much larger
than at the magic angle.
The results of Figs.~\ref{fig:Ds_vs_theta}(a) clearly show that $\Ds$ varies strongly with the twist angle,
and that, as a function of $\theta$, the dominant contribution to $\Ds$ can either be the conventional
one or the geometric. It is somewhat surprising, that even for twist angles as small as $1.00\degree$, corresponding to 
a bandwidth of the lowest energy bands of just 5~meV, the conventional contribution is larger than the geometric one. 

Figure~\ref{fig:Ds_vs_theta}~(b) shows the dependence of $T_c$ and $T_{\mathrm{KT}}$ on the twist angle.
We see that both $T_c$ and $T_{\mathrm{KT}}$ are maximum at the magic angle and decrease rapidly for $\theta$ larger than
the magic angle. 
The results of  Fig.~\ref{fig:Ds_vs_theta}~(b) suggests that it may be possible to tune $T_{\mathrm{KT}}$ by
tuning the twist angle. Taking into account finite size effects, this can change the nature of the normal-superconductor phase transition.

\captionsetup{justification=centerlast, indention=0cm, singlelinecheck=true}
\captionsetup{justification=centerlast, indention=0cm, singlelinecheck=true}

%
In summary, we have shown that in twisted bilayer graphene, despite the flatness of the low energy bands,
the superconducting weight $\Ds$ is finite and large enough to explain the experimental observation of superconducting behavior in these systems.
We find that the share of the geometric and conventional contributions to $\Ds$ depends on the twist angle:
at the magic angle the geometric contribution dominates, for angles slightly away from the magic angle the 
conventional contribution dominates. This qualitative difference is also reflected in the scaling
of $\Ds$ with $\mu$: at the magic angle $\Ds$ decreases with $|\mu|$, a somewhat surprising result due to the conventional contribution to $\Ds$ being almost
independent of $\mu$ whereas the geometric, large, contribution decreases with $|\mu|$ at the magic angle.
Away from the magic angle we find the more conventional behavior of $\Ds$ growing with $|\mu|$
as the conventional contribution, that grows with $|\mu|$, dominates.
This has the important implication that at the magic angle, by simply increasing $|\mu|$
twisted bilayer graphene can be tuned into a regime for which the Berezinzki-Kosterlits-Thouless transition
is significantly smaller than the BCS critical temperature. This result shows that twisted
bilayer graphene is an exceptional system in which the nature, BKT or BCS, of the superconducting
transition can be tuned and experimentally studied in unprecedented ways. 
We stress that the superfluid weight is one of the few observable signatures of the Berry phase in the Brillouin zone.
Measurements of the superfluid weight with the known experimental techniques~\cite{Hebard1980,Turneaure2000,Bert2011,Bert2012,Kapon2019} can directly test our predictions 
of its parametric dependencies.

\begin{acknowledgements}
{\em Acknowledgements.}
X.H. and E.R. acknowledge support NSF (CAREER Grant No. DMR-1455233) and ONR (Grant No. ONR- N00014-16-1-3158). ER also acknowledges support from ARO (Grant No. W911NF-18-1-0290). E.R. also thanks the Aspen Center for Physics, which is supported by National Science Foundation grant PHY-1607611, where part of this work was performed. The numerical calculations have been performed on computing facilities at William \& Mary which were provided by contributions from the NSF, the Commonwealth of Virginia Equipment Trust Fund, and ONR. T.H. was supported by the Foundation for Polish Science through the IRA Programme co-financed by EU within SG OP.
\end{acknowledgements}

{\it Note:} In the process of completing the manuscript, we became aware of a related recent preprint by  Aleksi Julku {\it et al.}~\cite{Julku2019}.


%

\bibliographystyle{apsrev4-1}
%



\setcounter{equation}{0}
\setcounter{figure}{0}
\setcounter{table}{0}
\renewcommand{\theequation}{S\arabic{equation}}
\renewcommand{\thefigure}{S\arabic{figure}}
\renewcommand{\thetable}{S\arabic{table}}
\renewcommand{\bibnumfmt}[1]{[S#1]}
\renewcommand{\citenumfont}[1]{S#1}
\newcommand{\bk}{\boldsymbol\kappa}

\section{Supplemental material}

\subsection{I. Solutions of the mean field gap equation}

To obtain an estimate of pairing parameters
we solve the linearized gap equation~\cite{Wu2018S}
$\Delta_{{\bf b}l}=\sum_{{\bf b}^{'} l^{'}}\chi_{{\bf b}{\bf b}^{'}}^{ll^{'}}
\Delta_{{\bf b}^{'} l^{'}}$,
where the pairing susceptibility is 
\begin{equation}
\chi_{{\bf b}{\bf b}^{'}}^{ll^{'}}=\frac{2g_0}{\mathcal{A}}\sum_{{\bf q}n_1 n_2}\mathcal{N}(\qq)[U({\bf q})_{{\bf b}l}]^*U({\bf q})_{{\bf b}^{'} l^{'}}.
\end{equation}
Here $\mathcal{N}(\qq)=\frac{1-n_F[\epsilon_{n_1}({\bf q})]-n_F[\epsilon_{n_2}(\bf q)]}{\epsilon_{n_1}({\bf q})+\epsilon_{n_2}({\bf q})-2\mu}$,  
$U({\bf q})_{{\bf b}l}=\langle u_{n_1}({\bf q})|e^{i{\bf b}\cdot{\bf r}}|
u_{n_2}({\bf q})\rangle_l$, $g_0$ is the pairing coupling constant due to electron-phonon interaction, $\mathcal{A}$ is the total area of the sample, $\epsilon_{n_{1,2}}({\bf q})$ and $u_{n_{1,2}}$ are eigenvalue and wavefunctions of the non-superconducting Hamiltonian $H_{TBL, \KK}(\kk)$ and $l(l^{'})$ represents the layer index. We keep the reciprocal basis vectors whose lengths are no larger than 
twice of $b\equiv|{\bf b}_1|$. 

When the temperature approaches $T_c$, the maximum eigenvalues of $\chi$ approach one. 
Correspondingly, the linearized gap equation becomes
\begin{equation}
\chi{\bf v}={\bf v},
\end{equation}
here ${\bf v}$ contains $\Delta_0$, $\Delta_b$ and higher order Fourier components of $\Delta({\bf r})$.
Because the higher order terms are small, we can neglect them and keep only $\Delta_0$ and $\Delta_b$.
The results are listed in table (\ref{tab:Delta}). 

\begin{table}[htb]
\centering
\caption{Results obtained from solving the mean-field gap equation.}
\begin{subtable}[t]{\linewidth}
\centering
\caption{Solutions at $\mu=-0.30$meV}
\begin{tabular}{||c|c|c|c|c||}
  \hline
  \hline
    $\theta(\degree)$ & 1.00 & 1.05 & 1.10  & 1.15 \\
  \hline
    $T_c(K)$ & 1.2119 & 1.6305 &  0.2340  &  0.0189 \\
  \hline
   $\Delta_0$ & 0.4193 & 0.4346  & 0.4465 & 0.4497 \\    
  \hline
   $\Delta_b$ & 0.2138 & 0.2118  &  0.2100  & 0.2104 \\
  \hline
 \end{tabular}
\end{subtable}
\hfill
\vspace{1mm}
\begin{subtable}[t]{\linewidth}
\centering
\caption{Solutions for $\theta=1.05^\circ$.}
\begin{tabular}{||c|c|c|c|c|c|c||}
  \hline
  \hline
    $\mu(meV)$ & 0 & -0.10  & -0.20  & -0.30  &  -0.40  &  -0.50\\
  \hline
    $T_c(K)$ & 1.7392 & 1.7330 & 1.7043 & 1.6305 & 1.4877 & 1.2455\\
  \hline
   $\Delta_0$ & 0.4346 & 0.4346 & 0.4346 & 0.4346 & 0.4346 & 0.4346\\
  \hline
   $\Delta_b$ & 0.2119 & 0.2119 & 0.2119 & 0.2118 & 0.2118 & 0.2118\\
  \hline
 \end{tabular}
\end{subtable}
\hfill
\vspace{1mm}
\begin{subtable}[t]{\linewidth}
\centering
\caption{Solutions for $\theta=1.00^\circ$.}
\begin{tabular}{||c|c|c|c|c|c|c||}
  \hline
  \hline
    $\mu(meV)$ & 0 & -0.10  & -0.20  & -0.30  &  -0.40  &  -0.50\\
  \hline
    $T_c(K)$ & 1.2086 & 1.2097 & 1.2219 & 1.2119 & 1.1561 & 1.0461\\
  \hline
   $\Delta_0$ & 0.4186 & 0.4188 & 0.4191 & 0.4193 & 0.4196 & 0.4197\\
  \hline
   $\Delta_b$ & 0.2138 & 0.2138 & 0.2138 & 0.2138 & 0.2138 & 0.2138\\
  \hline
 \end{tabular}
\end{subtable}

\label{tab:Delta}
\end{table}
Everywhere in the text we have asssumed $\Delta=1.764k_B T_c$ at zero temperature, where 1.764 is the prefactor for weak-coupling theory in metals. This assumption is made for simplicity because the actual prefactor in flat band systems depends on the details of the model.

\subsection{II. Calculation of the superfluid weight}

Our starting point is a BdG Hamiltonian for a singlet superconductor
\begin{equation}
H_{BdG}(\mathbf{p})=\begin{pmatrix}
H_0(\mathbf{p}) & \Delta(\mathbf{p}) \\
\Delta^\dag(\mathbf{p}) & -H_0^T(-\mathbf{p}) 
\end{pmatrix}.
\end{equation}
Here $H_0(\mathbf{p})$ is $n \times n$ normal state Hamiltonian and $\Delta(\mathbf{p})$ is the superconducting order parameter. 

We assume that the order parameter supports a gradient of phase $\phi=\mathbf{k}\cdot \mathbf{r}$
\begin{equation}
H_{BdG}(\mathbf{r})=\begin{pmatrix}
H_0(-i \nabla) & e^{i \frac{\mathbf{k}}{2}\cdot{\mathbf{r}}} \Delta(-i \nabla) e^{i \frac{\mathbf{k}}{2}\cdot{\mathbf{r}}} \\
e^{-i \frac{\mathbf{k}}{2}\cdot{\mathbf{r}}} \Delta^\dag(-i \nabla) e^{-i \frac{\mathbf{k}}{2}\cdot{\mathbf{r}}} & -H_0^T(+i\nabla)
\end{pmatrix}. \label{order-parameter-gradient}
\end{equation}
leading to a current 
\begin{equation}
\mathbf{j}=\frac{2 e D_s}{\hbar} \nabla \phi, \label{current-gradient}
\end{equation}
where $D_s$ is the superfluid weight. We choose the units so that $e=1$ and $\hbar = 1$ everywhere. The BKT transition temperature is given by
\begin{equation}
T_{\rm KT}=\frac{\pi}{2} D_s(T_{\rm KT}). \label{TKTstandard}
\end{equation}

We point out that in the literature several different conventions are used. In particular, one often uses $e^{i\phi(\mathbf{r})/2} \Delta(-i \nabla) e^{i\phi(\mathbf{r})/2}\to e^{i\phi(\mathbf{r})} \Delta(-i \nabla) e^{i\phi(\mathbf{r})}$ in Eq.~(\ref{order-parameter-gradient}) and $2e \to e$ in Eq.~(\ref{current-gradient}). With this convention $D_s$ becomes 4 times larger and $\pi/2$ is replaced by $\pi/8$, but this of course does not affect the predictions for $T_{\rm KT}$.

To diagonalize the Hamiltonian (\ref{order-parameter-gradient}) we use an ansatz 
\begin{equation}
\psi(\mathbf{r})=\begin{pmatrix}
a(\mathbf{p}) e^{i\mathbf{p} \cdot{\mathbf{r}}} e^{i \frac{\mathbf{k}}{2}\cdot \mathbf{r}}\\
b(\mathbf{p}) e^{i\mathbf{p} \cdot{\mathbf{r}}} e^{-i \frac{\mathbf{k}}{2}\cdot \mathbf{r}}
\end{pmatrix}.
\end{equation}
This way we find that $a(\mathbf{p})$,  $b(\mathbf{p})$ and corresponding eigenenergies $E(\mathbf{p})$ can be solved using the effective Hamiltonian
\begin{equation}
H_{\rm eff}(\mathbf{p}, \mathbf{k})=\begin{pmatrix}
H_0(\mathbf{p}+\frac{\mathbf{k}}{2}) & \Delta(\mathbf{p}) \\
\Delta^\dag(\mathbf{p}) & -H_0^T(-\mathbf{p}+\frac{\mathbf{k}}{2}) 
\end{pmatrix}.
\end{equation}

The operator for quasiparticle velocity in the basis described by the coefficients $a(\mathbf{p})$, $b(\mathbf{p})$ is given by 
\begin{equation}
v_\mu(\mathbf{p}, \mathbf{k})=\frac{\partial H_{\rm eff}(\mathbf{p}, \mathbf{k}, \Delta=0)}{\partial p_\mu}.
\end{equation}
The current density is given by
\begin{equation}
\mathbf{j}=\frac{1}{L^2}\sum_{\mathbf{p}, n} \langle n, \mathbf{p}, \mathbf{k}| \gamma_z \mathbf{v}(\mathbf{p}, \mathbf{k}) | n, \mathbf{p}, \mathbf{k} \rangle n_F(E_n(\mathbf{p}, \mathbf{k})),
\end{equation}
where $\gamma_z$ is the Pauli matrix in the Nambu space,  $| n, \mathbf{p}, \mathbf{k} \rangle$ are the eigenstates and $E_n(\mathbf{p}, \mathbf{k})$ the corresponding eigenenergies of Hamiltonian $H_{\rm eff}(\mathbf{p}, \mathbf{k})$.

The Hamiltonian (in Nambu space $C_{\mathbf{p}, \mathbf{k}}^\dag$)  can be written as 
\begin{equation}
\hat{H}=\sum_{\mathbf{p}} C_{\mathbf{p}, \mathbf{k}}^\dag H_{\rm eff}(\mathbf{p}, \mathbf{k}) C_{\mathbf{p}, \mathbf{k}}=\hat{H}^0+\hat{H}^1,
\end{equation}
where
\begin{equation}
\hat{H}^0=\sum_{\mathbf{p}} C_{\mathbf{p}, \mathbf{k}}^\dag  H_{\rm eff}(\mathbf{p}, \mathbf{0}) C_{\mathbf{p}, \mathbf{k}},
\end{equation}
and
\begin{equation}
  \hat{H}^1=\sum_{\mathbf{p}} C_{\mathbf{p}, \mathbf{k}}^\dag \gamma_z \frac{\mathbf{k}}{2} \cdot \mathbf{v}(\mathbf{p}, \mathbf{0}) C_{\mathbf{p}, \mathbf{k}}
\end{equation}

The current density operator can be written as
\begin{equation}
\hat{\mathbf{j}}=\frac{1}{L^2}\sum_{\mathbf{p}} C_{\mathbf{p}, \mathbf{k}}^\dag \gamma_z \mathbf{v}(\mathbf{p}, \mathbf{k}) C_{\mathbf{p}, \mathbf{k}}=\hat{\mathbf{j}}^0+\hat{\mathbf{j}}^1,
\end{equation}
where
\begin{equation}
\hat{\mathbf{j}}^0=\frac{1}{L^2}\sum_{\mathbf{p}} C_{\mathbf{p}, \mathbf{k}}^\dag \gamma_z \mathbf{v}(\mathbf{p}, 0) C_{\mathbf{p}, \mathbf{k}}
\end{equation}
and
\begin{equation}
\hat{\mathbf{j}}^1=\frac{1}{L^2}\sum_{\mathbf{p}} C_{\mathbf{p}, \mathbf{k}}^\dag \gamma_z \mathbf{k} \cdot \big[\nabla_{\mathbf{k}}\mathbf{v}(\mathbf{p}, \mathbf{k})\big]_{\mathbf{k}=\mathbf{0}} C_{\mathbf{p}, \mathbf{k}}.
\end{equation}

We assume that the current flows parallel to the gradient of phase $\mathbf{k}=k \mathbf{e}_\mu$, $\langle \hat{\mathbf{j}} \rangle =j \mathbf{e}_\mu$ and by using linear response theory to obtain 
\begin{eqnarray}
j&=&  \frac{k}{2L^2} \bigg\{ \sum_{\mathbf{p}, i} \langle \psi_i(\mathbf{p}) | \gamma_z T_{\mu\mu}(\mathbf{p}) | \psi_i(\mathbf{p}) \rangle n_F(E_i(\mathbf{p})) \nonumber \\
&& \hspace{-0.7cm}+ \sum_{\mathbf{p}, i, j} \frac{n_F(E_i(\mathbf{p}))-n_F(E_j(\mathbf{p}))}{E_i(\mathbf{p})-E_j(\mathbf{p})}  |\langle \psi_i(\mathbf{p}) | \gamma_z v_\mu(\mathbf{p}) | \psi_j(\mathbf{p}) \rangle |^2\bigg\}, 
\nonumber
\end{eqnarray} 
where 
\begin{equation}
T_{\mu \nu}(\mathbf{p})=\gamma_z \frac{\partial}{\partial p_\mu} v_{\nu}(\mathbf{p}),
\end{equation}
$E_i(\mathbf{p})$ and $\psi_i(\mathbf{p})$ are the eigenenergies and eigenstates of $H_{BdG}(\mathbf{p})$. We have also used a shorthand notation $v_\mu(\mathbf{p})=v_\mu(\mathbf{p}, \mathbf{k}=0)$. Therefore, we can identify 
\begin{eqnarray}
D_s &=& \frac{1}{4L^2} \bigg\{ \sum_{\mathbf{p}, i} \langle \psi_i(\mathbf{p}) | \gamma_z T_{\mu\mu}(\mathbf{p}) | \psi_i(\mathbf{p}) \rangle n_F(E_i(\mathbf{p})) \nonumber \\
&& \hspace{-0.95cm}+ \sum_{\mathbf{p}, i, j} \frac{n_F(E_i(\mathbf{p}))-n_F(E_j(\mathbf{p}))}{E_i(\mathbf{p})-E_j(\mathbf{p})}  |\langle \psi_i(\mathbf{p}) | \gamma_z v_\mu(\mathbf{p}) | \psi_j(\mathbf{p}) \rangle |^2\bigg\}, 
\nonumber
\end{eqnarray}
Additionally we can simplify the expression using
\begin{eqnarray}
&&\sum_{\mathbf{p}, i} \langle \psi_i(\mathbf{p}) | \tau_z T_{\mu\mu}(\mathbf{p}) | \psi_i(\mathbf{p}) \rangle n_F(E_i(\mathbf{p}))\nonumber\\&=&\frac{1}{\beta} \sum_{\omega_n} \sum_{\mathbf{p}} {\rm Tr}\big[G(\omega_n, \mathbf{p}) \tau_z T_{\mu \mu}(\mathbf{p}) \big] \nonumber\\
&=&-\frac{1}{\beta} \sum_{\omega_n} \sum_{\mathbf{p}} {\rm Tr}\bigg[\frac{\partial G(\omega_n, \mathbf{p})}{\partial p_\mu}  v_{\mu}(\mathbf{p}) \bigg] \nonumber \\
&=&-\frac{1}{\beta} \sum_{\omega_n} \sum_{\mathbf{p}} {\rm Tr}\bigg[G(\omega_n, \mathbf{p}) \frac{\partial H_{BdG}(\mathbf{p})}{\partial p_\mu} G(\omega_n, \mathbf{p}) v_{\mu}(\mathbf{p}) \bigg] \nonumber\\
&=&-\frac{1}{\beta} \sum_{\omega_n} \sum_{\mathbf{p}, i, j} \frac{1}{i \omega_n-E_i(\mathbf{p})} \frac{1}{i \omega_n-E_j(\mathbf{p})} \nonumber\\ && \hspace{1cm} \times \langle \psi_i(\mathbf{p}) | \frac{\partial H_{BdG}(\mathbf{p})}{\partial p_\mu} |\psi_j(\mathbf{p})\rangle \langle \psi_j(\mathbf{p}) |  v_{\mu}(\mathbf{p})  |\psi_i(\mathbf{p})\rangle\nonumber\\
&=&-\sum_{\mathbf{p}, i, j} \frac{n_F(E_i(\mathbf{p}))-n_F(E_j(\mathbf{p}))}{E_i(\mathbf{p})-E_j(\mathbf{p})} \nonumber\\ && \hspace{0.5cm} \times \langle \psi_i(\mathbf{p}) | \frac{\partial H_{BdG}(\mathbf{p})}{\partial p_\mu} |\psi_j(\mathbf{p})\rangle \langle \psi_j(\mathbf{p}) | v_{\mu}(\mathbf{p})  |\psi_i(\mathbf{p})\rangle, \nonumber
\end{eqnarray}
where $G(\omega_n,\mathbf{p})=(i \omega_n-H_{BdG}(\mathbf{p}))^{-1}$ and $\omega_n$ are Matsubara frequencies.
By assuming also that $\Delta(\mathbf{p})$ is independent of momentum $\mathbf{p}$ we obtain
\begin{equation}
v_\mu(\mathbf{p})=\frac{\partial H_{BdG}(\mathbf{p})}{\partial p_\mu}
\end{equation}
and arrive at the expression used in the main text.

\captionsetup{justification=centerlast, indention=0cm, singlelinecheck=true}
\begin{figure}[htbp]
\begin{center}
\includegraphics[width=\columnwidth]{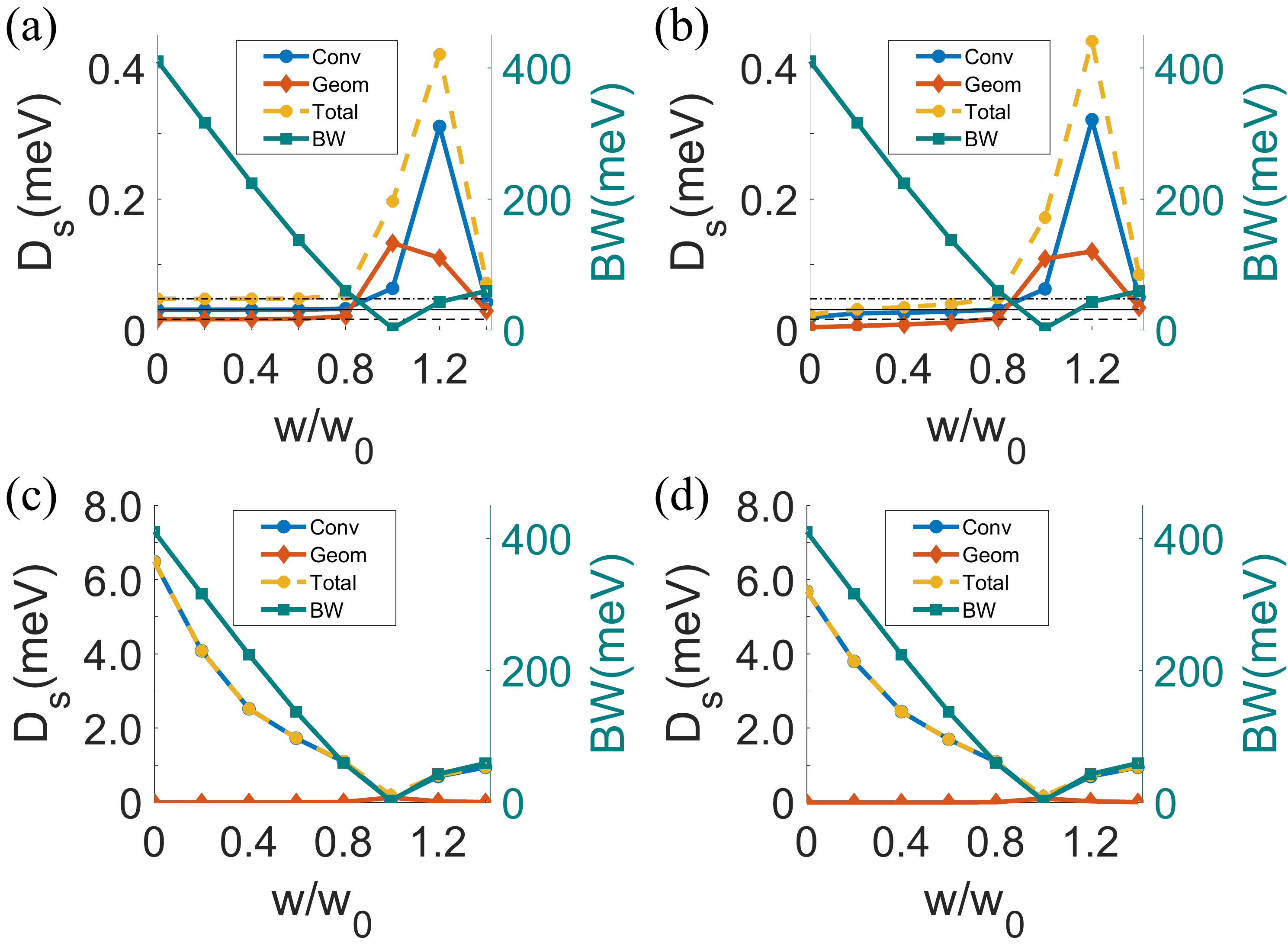}
\caption{The dependence of superfluid weight and bandwidth (BW) on the interlayer hopping $w$ at $\theta=1.05^\circ$. 
         Here $w_0=118$meV. 
         (a) Fixed chemical potential, and $\mu=-0.30$meV, and uniform pairing with $\Delta_0=1.764k_B T_c$.
         The straight lines show the analytical results at $w/w_0=0$. 
         (b) Fixed chemical potential, and $\mu=-0.30$meV, and spatially dependent pairing $\Delta(\mathbf{r})$ with parameters from Table \ref{tab:Delta}.
         (c) Fixed electrons' density and uniform pairing with $\Delta_0=1.764k_B T_c$. The density is fixed to the value corresponding to $\mu=-0.30$meV
             and $w/w_0=1$.
         (d) Same as (c) but with spatially dependent pairing $\Delta(\mathbf{r})$ with parameters from Table \ref{tab:Delta}.   \hfill
\label{fig:Ds_vs_w}}
\end{center}
\end{figure}

For single isolated parabolic band at zero temperature the expressions above give $D_s \propto n/m^*$, where $n$ is the density and $m^*$ is the effective mass of the electrons. In this calculation one needs to take into account that the density of electrons within the band is constant, and therefore when the superconducting transition occurs the chemical potential is renormalized. In particular, in the limit of small density and large $m^*$ the chemical potential is renormalized well below the band. This can be understood as a BCS to BEC crossover driven by decreasing $n/(m^* \Delta)$. In a semimetal where both electron- and hole-like bands are present, such as TBLG, similar renormalization of the chemical potential below the conduction band (above the valence band) cannot occur because it would lead to appearance of large density of holes (electrons) in the valence band (conduction band). The densities in each band are not conserved in the transition separately (only the total density is conserved) and therefore the conventional contribution to the superfluid weight is  necessarily much larger at small densities than expected for isolated bands.

In this manuscript we always consider a system consisting of two valleys $\KK$ and $\KK'$ and calculate $D_s$ only for a single valley. The valley degeneracy gives an additional factor of 2, so that $\pi/2$ in Eq.~(\ref{TKTstandard}) has to be replaced with $\pi$.

To check our numerical codes we have calculated the superfluid weight for a single valley of TBLG  numerically as a function of the effective interlayer tunneling amplitude $w$ in the case of uniform $s$-wave pairing [see Fig.\ref{fig:Ds_vs_w} (a)]. Analytically, for $w=0$ one obtains 
\begin{equation}
D_s=\frac{\Delta}{4\pi} \bigg[\sqrt{1+\frac{\mu^2}{\Delta^2}}+\frac{\Delta}{|\mu|}\ln\bigg(\frac{|\mu|}{\Delta}+\sqrt{1+\frac{\mu^2}{\Delta^2}}\bigg) \bigg],
\end{equation}
where the first term is the conventional superfluid weight and the second term is the geometric\cite{Liang2017S}. The numerics demonstrate that the superfluid weight quickly approaches the analytical result when the coupling between layers is decreased. By increasing $w$ we find that $D_s/\Delta$ first increases and then decreases. Qualitatively similar results are obtained also in the case of spatially dependent $\Delta(\mathbf{r})$ with parameters from Table \ref{tab:Delta}. However, in this calculation we have not self-consistently taken into account the influence of $w$ on the superconducting order parameter. Experimentally the effect of $w$ can be studied by applying pressure.

\subsection{III. The tight binding model of twisted bilayer graphene}

Following the approach described in Ref.~\onlinecite{Bistritzer2011aS,Wu2018S}, we can construct the tight
binding model of TBLG. In TBLG, the conservation of the crystal momentum requires $\kk_b = \kk_t + (\GG_t-\GG_b)$ where $\GG_i$ is the reciprocal lattice wave-vector in layer $i$, with $i=t,b$ representing the top or bottom layer. For small twist angles a fairly accurate description is obtained by just keeping the tunneling processes for which $|\kk_b - \kk_t|=|\GG_t-\GG_b|=2K\sin(\theta/2)$, where $K=4\pi/3a_0$. There are three vectors ${\bf Q}_i = \GG_t-\GG_b$ ($i=1,2,3)$ for which $Q=2K\sin(\theta/2)$ and so all the tunneling processes for which $|\kk_b - \kk_t|=Q$ are taken into account by keeping all the recursive tunneling processes on a honeycomb structure constructed in momentum space with nearest neighbor points connected by the vectors ${\bf Q}_i$. The honeycomb arrangement can be thought of as a triangular lattice. By keeping ${\bf b}=0,{\bf b}_1, {\bf b}_2$ we can write out the tight binding Hamiltonian around the $\KK$ point
\begin{widetext}
\begin{equation}
H_{TBL,{\bf K}}=\sum_{\bf k}\psi^\dagger(\kk)
\left(
\begin{array}{cc|cc|cc}
H_{\KK,+1}&T_0&&&&\\
T_0^\dagger&H_{\KK,-1}&T_1^\dagger&&T_{-1}^\dagger&\\
\hline
&T_1&H_{\KK+{\bf b}_1,+1}&T_0&&\\
&&T_0^\dagger&H_{\KK+{\bf b}_1,-1}&&\\
\hline
&T_{-1}&&&H_{\KK+{\bf b}_2,+1}&T_0\\
&&&&T_0^\dagger&H_{\KK+{\bf b}_2,-1}\\
\end{array}
\right)
\psi({\bf k}),
\end{equation}
where $\bk_{+1}=(0,0)$ and $\bk_{-1}=(0,Q)$, and the basis is
\begin{equation}
\psi(\kk)=(\phi_{\KK+\bk_{+1}+\kk},\phi_{\KK+\bk_{-1}+\kk},
\phi_{\KK+\bk_{+1}+{\bf b}_1+\kk},\phi_{\KK+\bk_{-1}+{\bf b}_1+\kk},
\phi_{\KK+\bk_{+1}+{\bf b}_2+\kk},\phi_{\KK+\bk_{-1}+{\bf b}_2+\kk})^T,
\end{equation}
\end{widetext}
with $\phi_{\kk}=(c_{\kk A},c_{\kk B})$. Here $c_{\kk A,B}$ is the electron 
annihilation operator with momentum $\kk$ at sublattice A,B. Similarly, 
We can include more ${\bf b}$ in this Hamiltonian.

\subsection{IV. Dependence of the superfluid weight on the number of bands}

\captionsetup{justification=centerlast, indention=0cm, singlelinecheck=true}
\begin{figure}[htbp]
  \includegraphics[width=\columnwidth]{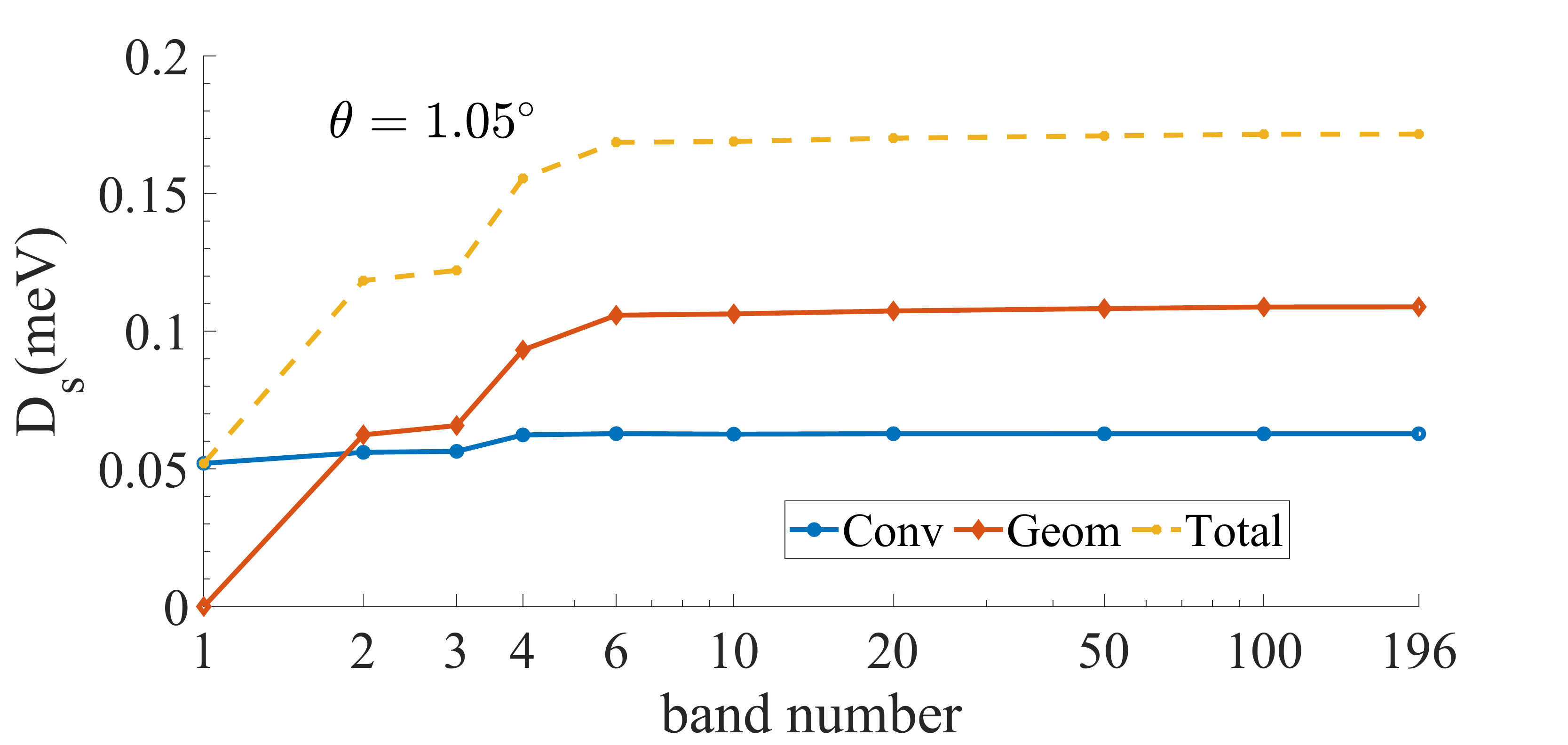}
  \caption{Dependence of superfluid weight on the number of non-superconducting bands included 
	  in the calculation. Here $\theta=1.05^\circ$ and $\mu=-0.30$meV. \hfill
  \label{fig:bdnb}}
\end{figure}

It is interesting to study how the superfluid weight depends on the number of bands that are included in the calculation.
Figure~\ref{fig:bdnb} shows the dependence of both the conventional and geometric part of $D^s$ on the number of bands.
We see that the conventional part depends only weakly on the number of bands, $n_{\rm bands}$, but that the geometric part 
depends very strongly on $n_{\rm bands}$. 
We see that keeping only the two nearly flat bands is not enough to get accurate estimates of the geometric contribution.
However, we find that when $n_{\rm bands}=10$, the $D^s_{\rm geom}$ is already very close (less than 2\% away) to the value 
obtained keeping as many as 196 bands. This seems consistent with recent results that suggest that a minimal model
for TBLG might require a minimum of 10 bands~\cite{Po2018aS}


%

\bibliographystyle{apsrev4-1}

\begin{thebibliography}{55}%
\makeatletter
\providecommand \@ifxundefined [1]{%
 \@ifx{#1\undefined}
}%
\providecommand \@ifnum [1]{%
 \ifnum #1\expandafter \@firstoftwo
 \else \expandafter \@secondoftwo
 \fi
}%
\providecommand \@ifx [1]{%
 \ifx #1\expandafter \@firstoftwo
 \else \expandafter \@secondoftwo
 \fi
}%
\providecommand \natexlab [1]{#1}%
\providecommand \enquote  [1]{``#1''}%
\providecommand \bibnamefont  [1]{#1}%
\providecommand \bibfnamefont [1]{#1}%
\providecommand \citenamefont [1]{#1}%
\providecommand \href@noop [0]{\@secondoftwo}%
\providecommand \href [0]{\begingroup \@sanitize@url \@href}%
\providecommand \@href[1]{\@@startlink{#1}\@@href}%
\providecommand \@@href[1]{\endgroup#1\@@endlink}%
\providecommand \@sanitize@url [0]{\catcode `\\12\catcode `\$12\catcode
  `\&12\catcode `\#12\catcode `\^12\catcode `\_12\catcode `\%12\relax}%
\providecommand \@@startlink[1]{}%
\providecommand \@@endlink[0]{}%
\providecommand \url  [0]{\begingroup\@sanitize@url \@url }%
\providecommand \@url [1]{\endgroup\@href {#1}{\urlprefix }}%
\providecommand \urlprefix  [0]{URL }%
\providecommand \Eprint [0]{\href }%
\providecommand \doibase [0]{http://dx.doi.org/}%
\providecommand \selectlanguage [0]{\@gobble}%
\providecommand \bibinfo  [0]{\@secondoftwo}%
\providecommand \bibfield  [0]{\@secondoftwo}%
\providecommand \translation [1]{[#1]}%
\providecommand \BibitemOpen [0]{}%
\providecommand \bibitemStop [0]{}%
\providecommand \bibitemNoStop [0]{.\EOS\space}%
\providecommand \EOS [0]{\spacefactor3000\relax}%
\providecommand \BibitemShut  [1]{\csname bibitem#1\endcsname}%
\let\auto@bib@innerbib\@empty
\bibitem [{\citenamefont {Li}\ \emph {et~al.}(2010)\citenamefont {Li},
  \citenamefont {Luican}, \citenamefont {{Lopes dos Santos}}, \citenamefont
  {{Castro Neto}}, \citenamefont {Reina}, \citenamefont {Kong},\ and\
  \citenamefont {Andrei}}]{Li2010}%
  \BibitemOpen
  \bibfield  {author} {\bibinfo {author} {\bibfnamefont {G.}~\bibnamefont
  {Li}}, \bibinfo {author} {\bibfnamefont {A.}~\bibnamefont {Luican}}, \bibinfo
  {author} {\bibfnamefont {J.~M.~B.}\ \bibnamefont {{Lopes dos Santos}}},
  \bibinfo {author} {\bibfnamefont {A.~H.}\ \bibnamefont {{Castro Neto}}},
  \bibinfo {author} {\bibfnamefont {A.}~\bibnamefont {Reina}}, \bibinfo
  {author} {\bibfnamefont {J.}~\bibnamefont {Kong}}, \ and\ \bibinfo {author}
  {\bibfnamefont {E.~Y.}\ \bibnamefont {Andrei}},\ }\href {\doibase
  10.1038/nphys1463} {\bibfield  {journal} {\bibinfo  {journal} {Nature
  Physics}\ }\textbf {\bibinfo {volume} {6}},\ \bibinfo {pages} {109} (\bibinfo
  {year} {2010})}\BibitemShut {NoStop}%
\bibitem [{\citenamefont {Geim}\ and\ \citenamefont
  {Grigorieva}(2013)}]{Geim2013}%
  \BibitemOpen
  \bibfield  {author} {\bibinfo {author} {\bibfnamefont {A.~K.}\ \bibnamefont
  {Geim}}\ and\ \bibinfo {author} {\bibfnamefont {I.~V.}\ \bibnamefont
  {Grigorieva}},\ }\href {\doibase 10.1038/nature12385} {\bibfield  {journal}
  {\bibinfo  {journal} {Nature}\ }\textbf {\bibinfo {volume} {499}},\ \bibinfo
  {pages} {419} (\bibinfo {year} {2013})}\BibitemShut {NoStop}%
\bibitem [{\citenamefont {Novoselov}\ \emph {et~al.}(2016)\citenamefont
  {Novoselov}, \citenamefont {Mishchenko}, \citenamefont {Carvalho},\ and\
  \citenamefont {Neto}}]{Neto2016}%
  \BibitemOpen
  \bibfield  {author} {\bibinfo {author} {\bibfnamefont {K.~S.}\ \bibnamefont
  {Novoselov}}, \bibinfo {author} {\bibfnamefont {A.}~\bibnamefont
  {Mishchenko}}, \bibinfo {author} {\bibfnamefont {A.}~\bibnamefont
  {Carvalho}}, \ and\ \bibinfo {author} {\bibfnamefont {A.~H.~C.}\ \bibnamefont
  {Neto}},\ }\href {\doibase 10.1126/science.aac9439} {\bibfield  {journal}
  {\bibinfo  {journal} {Science}\ }\textbf {\bibinfo {volume} {353}},\ \bibinfo
  {pages} {aac9439} (\bibinfo {year} {2016})}\BibitemShut {NoStop}%
\bibitem [{\citenamefont {Ajayan}\ \emph {et~al.}(2016)\citenamefont {Ajayan},
  \citenamefont {Kim},\ and\ \citenamefont {Banerjee}}]{Kim2016}%
  \BibitemOpen
  \bibfield  {author} {\bibinfo {author} {\bibfnamefont {P.}~\bibnamefont
  {Ajayan}}, \bibinfo {author} {\bibfnamefont {P.}~\bibnamefont {Kim}}, \ and\
  \bibinfo {author} {\bibfnamefont {K.}~\bibnamefont {Banerjee}},\ }\href
  {\doibase 10.1063/PT.3.3297} {\bibfield  {journal} {\bibinfo  {journal}
  {Physics Today}\ }\textbf {\bibinfo {volume} {69}},\ \bibinfo {pages} {No.9, 38}
  (\bibinfo {year} {2016})}\BibitemShut {NoStop}%
\bibitem [{\citenamefont {Lopes~dos Santos}\ \emph {et~al.}(2007)\citenamefont
  {Lopes~dos Santos}, \citenamefont {Peres},\ and\ \citenamefont
  {Castro~Neto}}]{dossantos2007}%
  \BibitemOpen
  \bibfield  {author} {\bibinfo {author} {\bibfnamefont {J.~M.~B.}\
  \bibnamefont {Lopes~dos Santos}}, \bibinfo {author} {\bibfnamefont
  {N.~M.~R.}\ \bibnamefont {Peres}}, \ and\ \bibinfo {author} {\bibfnamefont
  {A.~H.}\ \bibnamefont {Castro~Neto}},\ }\href {\doibase
  10.1103/PhysRevLett.99.256802} {\bibfield  {journal} {\bibinfo  {journal}
  {Phys. Rev. Lett.}\ }\textbf {\bibinfo {volume} {99}},\ \bibinfo {pages}
  {256802} (\bibinfo {year} {2007})}\BibitemShut {NoStop}%
\bibitem [{\citenamefont {Mele}(2010)}]{Mele2010}%
  \BibitemOpen
  \bibfield  {author} {\bibinfo {author} {\bibfnamefont {E.~J.}\ \bibnamefont
  {Mele}},\ }\href {\doibase 10.1103/PhysRevB.81.161405} {\bibfield  {journal}
  {\bibinfo  {journal} {Phys. Rev. B}\ }\textbf {\bibinfo {volume} {81}},\
  \bibinfo {pages} {161405(R)} (\bibinfo {year} {2010})}\BibitemShut {NoStop}%
\bibitem [{\citenamefont {Morell}\ \emph {et~al.}(2010)\citenamefont {Morell},
  \citenamefont {Correa}, \citenamefont {Vargas}, \citenamefont {Pacheco},\
  and\ \citenamefont {Barticevic}}]{Morell2010}%
  \BibitemOpen
  \bibfield  {author} {\bibinfo {author} {\bibfnamefont {E.~S.}\ \bibnamefont
  {Morell}}, \bibinfo {author} {\bibfnamefont {J.~D.}\ \bibnamefont {Correa}},
  \bibinfo {author} {\bibfnamefont {P.}~\bibnamefont {Vargas}}, \bibinfo
  {author} {\bibfnamefont {M.}~\bibnamefont {Pacheco}}, \ and\ \bibinfo
  {author} {\bibfnamefont {Z.}~\bibnamefont {Barticevic}},\ }\href {\doibase
  10.1103/PhysRevB.82.121407} {\bibfield  {journal} {\bibinfo  {journal} {Phys.
  Rev. B}\ }\textbf {\bibinfo {volume} {82}},\ \bibinfo {pages} {121407(R)}
  (\bibinfo {year} {2010})}\BibitemShut {NoStop}%
\bibitem [{\citenamefont {Bistritzer}\ and\ \citenamefont
  {MacDonald}(2011)}]{Bistritzer2011a}%
  \BibitemOpen
  \bibfield  {author} {\bibinfo {author} {\bibfnamefont {R.}~\bibnamefont
  {Bistritzer}}\ and\ \bibinfo {author} {\bibfnamefont {A.~H.}\ \bibnamefont
  {MacDonald}},\ }\href {\doibase 10.1073/pnas.1108174108} {\bibfield
  {journal} {\bibinfo  {journal} {Proc. Natl. Acad. Sci. U.S.A.}\ }\textbf
  {\bibinfo {volume} {108}},\ \bibinfo {pages} {12233} (\bibinfo {year}
  {2011})}\BibitemShut {NoStop}%
\bibitem [{\citenamefont {Trambly~de Laissardi\`ere}\ \emph
  {et~al.}(2012)\citenamefont {Trambly~de Laissardi\`ere}, \citenamefont
  {Mayou},\ and\ \citenamefont {Magaud}}]{Magaud2012}%
  \BibitemOpen
  \bibfield  {author} {\bibinfo {author} {\bibfnamefont {G.}~\bibnamefont
  {Trambly~de Laissardi\`ere}}, \bibinfo {author} {\bibfnamefont
  {D.}~\bibnamefont {Mayou}}, \ and\ \bibinfo {author} {\bibfnamefont
  {L.}~\bibnamefont {Magaud}},\ }\href {\doibase 10.1103/PhysRevB.86.125413}
  {\bibfield  {journal} {\bibinfo  {journal} {Phys. Rev. B}\ }\textbf {\bibinfo
  {volume} {86}},\ \bibinfo {pages} {125413} (\bibinfo {year}
  {2012})}\BibitemShut {NoStop}%
\bibitem [{\citenamefont {Song}\ \emph {et~al.}(2019)\citenamefont {Song},
  \citenamefont {Wang}, \citenamefont {Shi}, \citenamefont {Li}, \citenamefont
  {Fang},\ and\ \citenamefont {Bernevig}}]{Song2019}%
  \BibitemOpen
  \bibfield  {author} {\bibinfo {author} {\bibfnamefont {Z.}~\bibnamefont
  {Song}}, \bibinfo {author} {\bibfnamefont {Z.}~\bibnamefont {Wang}}, \bibinfo
  {author} {\bibfnamefont {W.}~\bibnamefont {Shi}}, \bibinfo {author}
  {\bibfnamefont {G.}~\bibnamefont {Li}}, \bibinfo {author} {\bibfnamefont
  {C.}~\bibnamefont {Fang}}, \ and\ \bibinfo {author} {\bibfnamefont {B.~A.}\
  \bibnamefont {Bernevig}},\ }\href {\doibase 10.1103/PhysRevLett.123.036401}
  {\bibfield  {journal} {\bibinfo  {journal} {Phys. Rev. Lett.}\ }\textbf
  {\bibinfo {volume} {123}},\ \bibinfo {pages} {36401} (\bibinfo {year}
  {2019})}\BibitemShut {NoStop}%
\bibitem [{\citenamefont {{Lian}}\ \emph {et~al.}(2018)\citenamefont {{Lian}},
  \citenamefont {{Xie}},\ and\ \citenamefont {{Bernevig}}}]{Bernevig2018}%
  \BibitemOpen
  \bibfield  {author} {\bibinfo {author} {\bibfnamefont {B.}~\bibnamefont
  {{Lian}}}, \bibinfo {author} {\bibfnamefont {F.}~\bibnamefont {{Xie}}}, \
  and\ \bibinfo {author} {\bibfnamefont {B.~A.}\ \bibnamefont {{Bernevig}}},\
  }\href@noop {} {\bibfield  {journal} {\bibinfo  {journal} {arXiv e-prints}\
  ,\ \bibinfo {eid} {arXiv:1811.11786}} (\bibinfo {year} {2018})},\ \Eprint
  {http://arxiv.org/abs/1811.11786} {arXiv:1811.11786 [cond-mat.mes-hall]}
  \BibitemShut {NoStop}%
\bibitem [{\citenamefont {Po}\ \emph {et~al.}(2019)\citenamefont {Po},
  \citenamefont {Zou}, \citenamefont {Senthil},\ and\ \citenamefont
  {Vishwanath}}]{Po2019}%
  \BibitemOpen
  \bibfield  {author} {\bibinfo {author} {\bibfnamefont {H.~C.}\ \bibnamefont
  {Po}}, \bibinfo {author} {\bibfnamefont {L.}~\bibnamefont {Zou}}, \bibinfo
  {author} {\bibfnamefont {T.}~\bibnamefont {Senthil}}, \ and\ \bibinfo
  {author} {\bibfnamefont {A.}~\bibnamefont {Vishwanath}},\ }\href {\doibase
  10.1103/PhysRevB.99.195455} {\bibfield  {journal} {\bibinfo  {journal} {Phys.
  Rev. B}\ }\textbf {\bibinfo {volume} {99}},\ \bibinfo {pages} {195455}
  (\bibinfo {year} {2019})}\BibitemShut {NoStop}%
\bibitem [{\citenamefont {{Liu}}\ \emph
  {et~al.}(2019{\natexlab{a}})\citenamefont {{Liu}}, \citenamefont {{Liu}},\
  and\ \citenamefont {{Dai}}}]{Dai2019}%
  \BibitemOpen
  \bibfield  {author} {\bibinfo {author} {\bibfnamefont {J.}~\bibnamefont
  {{Liu}}}, \bibinfo {author} {\bibfnamefont {J.}~\bibnamefont {{Liu}}}, \ and\
  \bibinfo {author} {\bibfnamefont {X.}~\bibnamefont {{Dai}}},\ }\href
  {\doibase 10.1103/PhysRevB.99.155415} {\bibfield  {journal} {\bibinfo
  {journal} {\prb}\ }\textbf {\bibinfo {volume} {99}},\ \bibinfo {eid} {155415}
  (\bibinfo {year} {2019}{\natexlab{a}})},\ \Eprint
  {http://arxiv.org/abs/1810.03103} {arXiv:1810.03103 [cond-mat.mes-hall]}
  \BibitemShut {NoStop}%
\bibitem [{\citenamefont {{Ahn}}\ \emph {et~al.}(2019)\citenamefont {{Ahn}},
  \citenamefont {{Park}},\ and\ \citenamefont {{Yang}}}]{Yang2019}%
  \BibitemOpen
  \bibfield  {author} {\bibinfo {author} {\bibfnamefont {J.}~\bibnamefont
  {{Ahn}}}, \bibinfo {author} {\bibfnamefont {S.}~\bibnamefont {{Park}}}, \
  and\ \bibinfo {author} {\bibfnamefont {B.-J.}\ \bibnamefont {{Yang}}},\
  }\href {\doibase 10.1103/PhysRevX.9.021013} {\bibfield  {journal} {\bibinfo
  {journal} {Phys. Rev. X}\ }\textbf {\bibinfo {volume} {9}},\ \bibinfo {eid}
  {021013} (\bibinfo {year} {2019})},\ \Eprint
  {http://arxiv.org/abs/1808.05375} {arXiv:1808.05375 [cond-mat.mes-hall]}
  \BibitemShut {NoStop}%
\bibitem [{\citenamefont {Kopnin}\ \emph {et~al.}(2011)\citenamefont {Kopnin},
  \citenamefont {Heikkil\"a},\ and\ \citenamefont {Volovik}}]{Kopnin2011}%
  \BibitemOpen
  \bibfield  {author} {\bibinfo {author} {\bibfnamefont {N.~B.}\ \bibnamefont
  {Kopnin}}, \bibinfo {author} {\bibfnamefont {T.~T.}\ \bibnamefont
  {Heikkil\"a}}, \ and\ \bibinfo {author} {\bibfnamefont {G.~E.}\ \bibnamefont
  {Volovik}},\ }\href {\doibase 10.1103/PhysRevB.83.220503} {\bibfield
  {journal} {\bibinfo  {journal} {Phys. Rev. B}\ }\textbf {\bibinfo {volume}
  {83}},\ \bibinfo {pages} {220503(R)} (\bibinfo {year} {2011})}\BibitemShut
  {NoStop}%
\bibitem [{\citenamefont {Ojaj\"arvi}\ \emph {et~al.}(2018)\citenamefont
  {Ojaj\"arvi}, \citenamefont {Hyart}, \citenamefont {Silaev},\ and\
  \citenamefont {Heikkil\"a}}]{Ojajarvi2018}%
  \BibitemOpen
  \bibfield  {author} {\bibinfo {author} {\bibfnamefont {R.}~\bibnamefont
  {Ojaj\"arvi}}, \bibinfo {author} {\bibfnamefont {T.}~\bibnamefont {Hyart}},
  \bibinfo {author} {\bibfnamefont {M.~A.}\ \bibnamefont {Silaev}}, \ and\
  \bibinfo {author} {\bibfnamefont {T.~T.}\ \bibnamefont {Heikkil\"a}},\ }\href
  {\doibase 10.1103/PhysRevB.98.054515} {\bibfield  {journal} {\bibinfo
  {journal} {Phys. Rev. B}\ }\textbf {\bibinfo {volume} {98}},\ \bibinfo
  {pages} {054515} (\bibinfo {year} {2018})}\BibitemShut {NoStop}%
\bibitem [{\citenamefont {Wu}\ \emph {et~al.}(2018)\citenamefont {Wu},
  \citenamefont {MacDonald},\ and\ \citenamefont {Martin}}]{Wu2018}%
  \BibitemOpen
  \bibfield  {author} {\bibinfo {author} {\bibfnamefont {F.}~\bibnamefont
  {Wu}}, \bibinfo {author} {\bibfnamefont {A.~H.}\ \bibnamefont {MacDonald}}, \
  and\ \bibinfo {author} {\bibfnamefont {I.}~\bibnamefont {Martin}},\ }\href
  {\doibase 10.1103/PhysRevLett.121.257001} {\bibfield  {journal} {\bibinfo
  {journal} {Phys. Rev. Lett.}\ }\textbf {\bibinfo {volume} {121}},\ \bibinfo
  {pages} {257001} (\bibinfo {year} {2018})},\ \Eprint
  {http://arxiv.org/abs/1805.08735} {arXiv:1805.08735} \BibitemShut {NoStop}%
\bibitem [{\citenamefont {Peltonen}\ \emph {et~al.}(2018)\citenamefont
  {Peltonen}, \citenamefont {Ojaj\"arvi},\ and\ \citenamefont
  {Heikkil\"a}}]{PeltonenPhysRevB.98.220504}%
  \BibitemOpen
  \bibfield  {author} {\bibinfo {author} {\bibfnamefont {T.~J.}\ \bibnamefont
  {Peltonen}}, \bibinfo {author} {\bibfnamefont {R.}~\bibnamefont
  {Ojaj\"arvi}}, \ and\ \bibinfo {author} {\bibfnamefont {T.~T.}\ \bibnamefont
  {Heikkil\"a}},\ }\href {\doibase 10.1103/PhysRevB.98.220504} {\bibfield
  {journal} {\bibinfo  {journal} {Phys. Rev. B}\ }\textbf {\bibinfo {volume}
  {98}},\ \bibinfo {pages} {220504(R)} (\bibinfo {year} {2018})}\BibitemShut
  {NoStop}%
\bibitem [{\citenamefont {Guo}\ \emph {et~al.}(2018)\citenamefont {Guo},
  \citenamefont {Zhu}, \citenamefont {Feng},\ and\ \citenamefont
  {Scalettar}}]{GuoPhysRevB.97.235453}%
  \BibitemOpen
  \bibfield  {author} {\bibinfo {author} {\bibfnamefont {H.}~\bibnamefont
  {Guo}}, \bibinfo {author} {\bibfnamefont {X.}~\bibnamefont {Zhu}}, \bibinfo
  {author} {\bibfnamefont {S.}~\bibnamefont {Feng}}, \ and\ \bibinfo {author}
  {\bibfnamefont {R.~T.}\ \bibnamefont {Scalettar}},\ }\href {\doibase
  10.1103/PhysRevB.97.235453} {\bibfield  {journal} {\bibinfo  {journal} {Phys.
  Rev. B}\ }\textbf {\bibinfo {volume} {97}},\ \bibinfo {pages} {235453}
  (\bibinfo {year} {2018})}\BibitemShut {NoStop}%
\bibitem [{\citenamefont {Liu}\ \emph {et~al.}(2018)\citenamefont {Liu},
  \citenamefont {Zhang}, \citenamefont {Chen},\ and\ \citenamefont
  {Yang}}]{LiuPhysRevLett.121.217001}%
  \BibitemOpen
  \bibfield  {author} {\bibinfo {author} {\bibfnamefont {C.-C.}\ \bibnamefont
  {Liu}}, \bibinfo {author} {\bibfnamefont {L.-D.}\ \bibnamefont {Zhang}},
  \bibinfo {author} {\bibfnamefont {W.-Q.}\ \bibnamefont {Chen}}, \ and\
  \bibinfo {author} {\bibfnamefont {F.}~\bibnamefont {Yang}},\ }\href {\doibase
  10.1103/PhysRevLett.121.217001} {\bibfield  {journal} {\bibinfo  {journal}
  {Phys. Rev. Lett.}\ }\textbf {\bibinfo {volume} {121}},\ \bibinfo {pages}
  {217001} (\bibinfo {year} {2018})}\BibitemShut {NoStop}%
\bibitem [{\citenamefont {{You}}\ and\ \citenamefont
  {{Vishwanath}}(2019)}]{Ashvin2019}%
  \BibitemOpen
  \bibfield  {author} {\bibinfo {author} {\bibfnamefont {Y.-Z.}\ \bibnamefont
  {{You}}}\ and\ \bibinfo {author} {\bibfnamefont {A.}~\bibnamefont
  {{Vishwanath}}},\ }\href {\doibase 10.1038/s41535-019-0153-4} {\bibfield
  {journal} {\bibinfo  {journal} {npj Quantum Materials}\ }\textbf {\bibinfo
  {volume} {4}},\ \bibinfo {eid} {16} (\bibinfo {year} {2019})},\ \Eprint
  {http://arxiv.org/abs/1805.06867} {arXiv:1805.06867 [cond-mat.str-el]}
  \BibitemShut {NoStop}%
\bibitem [{\citenamefont {Zhang}\ and\ \citenamefont
  {Senthil}(2019)}]{Senthil2018}%
  \BibitemOpen
  \bibfield  {author} {\bibinfo {author} {\bibfnamefont {Y.-H.}\ \bibnamefont
  {Zhang}}\ and\ \bibinfo {author} {\bibfnamefont {T.}~\bibnamefont
  {Senthil}},\ }\href {\doibase 10.1103/PhysRevB.99.205150} {\bibfield
  {journal} {\bibinfo  {journal} {Phys. Rev. B}\ }\textbf {\bibinfo {volume}
  {99}},\ \bibinfo {pages} {205150} (\bibinfo {year} {2019})}\BibitemShut
  {NoStop}%
\bibitem [{\citenamefont {Tang}\ \emph {et~al.}(2019)\citenamefont {Tang},
  \citenamefont {Yang}, \citenamefont {Wang}, \citenamefont {Zhang},\ and\
  \citenamefont {Wang}}]{TangPhysRevB.99.094521}%
  \BibitemOpen
  \bibfield  {author} {\bibinfo {author} {\bibfnamefont {Q.-K.}\ \bibnamefont
  {Tang}}, \bibinfo {author} {\bibfnamefont {L.}~\bibnamefont {Yang}}, \bibinfo
  {author} {\bibfnamefont {D.}~\bibnamefont {Wang}}, \bibinfo {author}
  {\bibfnamefont {F.-C.}\ \bibnamefont {Zhang}}, \ and\ \bibinfo {author}
  {\bibfnamefont {Q.-H.}\ \bibnamefont {Wang}},\ }\href {\doibase
  10.1103/PhysRevB.99.094521} {\bibfield  {journal} {\bibinfo  {journal} {Phys.
  Rev. B}\ }\textbf {\bibinfo {volume} {99}},\ \bibinfo {pages} {094521}
  (\bibinfo {year} {2019})}\BibitemShut {NoStop}%
\bibitem [{\citenamefont {Roy}\ and\ \citenamefont {Juri\ifmmode \check{c}\else
  \v{c}\fi{}i\ifmmode~\acute{c}\else
  \'{c}\fi{}}(2019)}]{RoyPhysRevB.99.121407}%
  \BibitemOpen
  \bibfield  {author} {\bibinfo {author} {\bibfnamefont {B.}~\bibnamefont
  {Roy}}\ and\ \bibinfo {author} {\bibfnamefont {V.}~\bibnamefont {Juri\ifmmode
  \check{c}\else \v{c}\fi{}i\ifmmode~\acute{c}\else \'{c}\fi{}}},\ }\href
  {\doibase 10.1103/PhysRevB.99.121407} {\bibfield  {journal} {\bibinfo
  {journal} {Phys. Rev. B}\ }\textbf {\bibinfo {volume} {99}},\ \bibinfo
  {pages} {121407(R)} (\bibinfo {year} {2019})}\BibitemShut {NoStop}%
\bibitem [{\citenamefont {{Kozii}}\ \emph {et~al.}(2019)\citenamefont
  {{Kozii}}, \citenamefont {{Isobe}}, \citenamefont {{Venderbos}},\ and\
  \citenamefont {{Fu}}}]{Fu2019}%
  \BibitemOpen
  \bibfield  {author} {\bibinfo {author} {\bibfnamefont {V.}~\bibnamefont
  {{Kozii}}}, \bibinfo {author} {\bibfnamefont {H.}~\bibnamefont {{Isobe}}},
  \bibinfo {author} {\bibfnamefont {J.~W.~F.}\ \bibnamefont {{Venderbos}}}, \
  and\ \bibinfo {author} {\bibfnamefont {L.}~\bibnamefont {{Fu}}},\ }\href
  {\doibase 10.1103/PhysRevB.99.144507} {\bibfield  {journal} {\bibinfo
  {journal} {\prb}\ }\textbf {\bibinfo {volume} {99}},\ \bibinfo {eid} {144507}
  (\bibinfo {year} {2019})},\ \Eprint {http://arxiv.org/abs/1810.04159}
  {arXiv:1810.04159 [cond-mat.supr-con]} \BibitemShut {NoStop}%
\bibitem [{\citenamefont {Wu}(2019)}]{WuPhysRevB.99.195114}%
  \BibitemOpen
  \bibfield  {author} {\bibinfo {author} {\bibfnamefont {F.}~\bibnamefont
  {Wu}},\ }\href {\doibase 10.1103/PhysRevB.99.195114} {\bibfield  {journal}
  {\bibinfo  {journal} {Phys. Rev. B}\ }\textbf {\bibinfo {volume} {99}},\
  \bibinfo {pages} {195114} (\bibinfo {year} {2019})}\BibitemShut {NoStop}%
\bibitem [{\citenamefont {{Wu}}\ \emph {et~al.}(2019)\citenamefont {{Wu}},
  \citenamefont {{Hwang}},\ and\ \citenamefont {{Das Sarma}}}]{Wu2019}%
  \BibitemOpen
  \bibfield  {author} {\bibinfo {author} {\bibfnamefont {F.}~\bibnamefont
  {{Wu}}}, \bibinfo {author} {\bibfnamefont {E.}~\bibnamefont {{Hwang}}}, \
  and\ \bibinfo {author} {\bibfnamefont {S.}~\bibnamefont {{Das Sarma}}},\
  }\href {\doibase 10.1103/PhysRevB.99.165112} {\bibfield  {journal} {\bibinfo
  {journal} {\prb}\ }\textbf {\bibinfo {volume} {99}},\ \bibinfo {eid} {165112}
  (\bibinfo {year} {2019})},\ \Eprint {http://arxiv.org/abs/1811.04920}
  {arXiv:1811.04920 [cond-mat.mes-hall]} \BibitemShut {NoStop}%
\bibitem [{\citenamefont {Kim}\ \emph {et~al.}(2017)\citenamefont {Kim},
  \citenamefont {Dasilva}, \citenamefont {Huang}, \citenamefont {Fallahazad},
  \citenamefont {Larentis}, \citenamefont {Taniguchi}, \citenamefont
  {Watanabe}, \citenamefont {LeRoy}, \citenamefont {MacDonald}, ,\ and\
  \citenamefont {Tutuc}}]{Kim2017}%
  \BibitemOpen
  \bibfield  {author} {\bibinfo {author} {\bibfnamefont {K.}~\bibnamefont
  {Kim}}, \bibinfo {author} {\bibfnamefont {A.}~\bibnamefont {Dasilva}},
  \bibinfo {author} {\bibfnamefont {S.}~\bibnamefont {Huang}}, \bibinfo
  {author} {\bibfnamefont {B.}~\bibnamefont {Fallahazad}}, \bibinfo {author}
  {\bibfnamefont {S.}~\bibnamefont {Larentis}}, \bibinfo {author}
  {\bibfnamefont {T.}~\bibnamefont {Taniguchi}}, \bibinfo {author}
  {\bibfnamefont {K.}~\bibnamefont {Watanabe}}, \bibinfo {author}
  {\bibfnamefont {B.~J.}\ \bibnamefont {LeRoy}}, \bibinfo {author}
  {\bibfnamefont {A.~H.}\ \bibnamefont {MacDonald}}, , \ and\ \bibinfo {author}
  {\bibfnamefont {E.}~\bibnamefont {Tutuc}},\ }\href {\doibase
  10.1073/pnas.1620140114} {\bibfield  {journal} {\bibinfo  {journal} {Proc.
  Natl. Acad. Sci. U.S.A.}\ }\textbf {\bibinfo {volume} {114}},\ \bibinfo
  {pages} {3364} (\bibinfo {year} {2017})}\BibitemShut {NoStop}%
\bibitem [{\citenamefont {Cao}\ \emph {et~al.}(2018{\natexlab{a}})\citenamefont
  {Cao}, \citenamefont {Fatemi}, \citenamefont {Fang}, \citenamefont
  {Watanabe}, \citenamefont {Taniguchi}, \citenamefont {Kaxiras},\ and\
  \citenamefont {Jarillo-Herrero}}]{Cao2018}%
  \BibitemOpen
  \bibfield  {author} {\bibinfo {author} {\bibfnamefont {Y.}~\bibnamefont
  {Cao}}, \bibinfo {author} {\bibfnamefont {V.}~\bibnamefont {Fatemi}},
  \bibinfo {author} {\bibfnamefont {S.}~\bibnamefont {Fang}}, \bibinfo {author}
  {\bibfnamefont {K.}~\bibnamefont {Watanabe}}, \bibinfo {author}
  {\bibfnamefont {T.}~\bibnamefont {Taniguchi}}, \bibinfo {author}
  {\bibfnamefont {E.}~\bibnamefont {Kaxiras}}, \ and\ \bibinfo {author}
  {\bibfnamefont {P.}~\bibnamefont {Jarillo-Herrero}},\ }\href {\doibase
  10.1038/nature26160} {\bibfield  {journal} {\bibinfo  {journal} {Nature}\
  }\textbf {\bibinfo {volume} {556}},\ \bibinfo {pages} {43} (\bibinfo {year}
  {2018}{\natexlab{a}})}\BibitemShut {NoStop}%
\bibitem [{\citenamefont {Cao}\ \emph {et~al.}(2018{\natexlab{b}})\citenamefont
  {Cao}, \citenamefont {Fatemi}, \citenamefont {Demir}, \citenamefont {Fang},
  \citenamefont {Tomarken}, \citenamefont {Luo}, \citenamefont
  {Sanchez-Yamagishi}, \citenamefont {Watanabe}, \citenamefont {Taniguchi},
  \citenamefont {Kaxiras}, \citenamefont {Ashoori},\ and\ \citenamefont
  {Jarillo-Herrero}}]{Cao2018b}%
  \BibitemOpen
  \bibfield  {author} {\bibinfo {author} {\bibfnamefont {Y.}~\bibnamefont
  {Cao}}, \bibinfo {author} {\bibfnamefont {V.}~\bibnamefont {Fatemi}},
  \bibinfo {author} {\bibfnamefont {A.}~\bibnamefont {Demir}}, \bibinfo
  {author} {\bibfnamefont {S.}~\bibnamefont {Fang}}, \bibinfo {author}
  {\bibfnamefont {S.~L.}\ \bibnamefont {Tomarken}}, \bibinfo {author}
  {\bibfnamefont {J.~Y.}\ \bibnamefont {Luo}}, \bibinfo {author} {\bibfnamefont
  {J.~D.}\ \bibnamefont {Sanchez-Yamagishi}}, \bibinfo {author} {\bibfnamefont
  {K.}~\bibnamefont {Watanabe}}, \bibinfo {author} {\bibfnamefont
  {T.}~\bibnamefont {Taniguchi}}, \bibinfo {author} {\bibfnamefont
  {E.}~\bibnamefont {Kaxiras}}, \bibinfo {author} {\bibfnamefont {R.~C.}\
  \bibnamefont {Ashoori}}, \ and\ \bibinfo {author} {\bibfnamefont
  {P.}~\bibnamefont {Jarillo-Herrero}},\ }\href {\doibase 10.1038/nature26154}
  {\bibfield  {journal} {\bibinfo  {journal} {Nature}\ }\textbf {\bibinfo
  {volume} {556}},\ \bibinfo {pages} {80} (\bibinfo {year}
  {2018}{\natexlab{b}})}\BibitemShut {NoStop}%
\bibitem [{\citenamefont {Yankowitz}\ \emph {et~al.}(2019)\citenamefont
  {Yankowitz}, \citenamefont {Chen}, \citenamefont {Polshyn}, \citenamefont
  {Zhang}, \citenamefont {Watanabe}, \citenamefont {Taniguchi}, \citenamefont
  {Graf}, \citenamefont {Young},\ and\ \citenamefont {Dean}}]{Yankowitz2019}%
  \BibitemOpen
  \bibfield  {author} {\bibinfo {author} {\bibfnamefont {M.}~\bibnamefont
  {Yankowitz}}, \bibinfo {author} {\bibfnamefont {S.}~\bibnamefont {Chen}},
  \bibinfo {author} {\bibfnamefont {H.}~\bibnamefont {Polshyn}}, \bibinfo
  {author} {\bibfnamefont {Y.}~\bibnamefont {Zhang}}, \bibinfo {author}
  {\bibfnamefont {K.}~\bibnamefont {Watanabe}}, \bibinfo {author}
  {\bibfnamefont {T.}~\bibnamefont {Taniguchi}}, \bibinfo {author}
  {\bibfnamefont {D.}~\bibnamefont {Graf}}, \bibinfo {author} {\bibfnamefont
  {A.~F.}\ \bibnamefont {Young}}, \ and\ \bibinfo {author} {\bibfnamefont
  {C.~R.}\ \bibnamefont {Dean}},\ }\href {\doibase 10.1126/science.aav1910}
  {\bibfield  {journal} {\bibinfo  {journal} {Science}\ } \textbf {\bibinfo
  {volume} {363}},\ \bibinfo {pages} {1059} (\bibinfo {year}
  {2019}),\ 10.1126/science.aav1910}\BibitemShut {NoStop}%
\bibitem [{\citenamefont {Chen}\ \emph {et~al.}(2019)\citenamefont {Chen},
  \citenamefont {Sharpe}, \citenamefont {Gallagher}, \citenamefont {Rosen},
  \citenamefont {Fox}, \citenamefont {Jiang}, \citenamefont {Lyu},
  \citenamefont {Li}, \citenamefont {Watanabe}, \citenamefont {Taniguchi},
  \citenamefont {Jung}, \citenamefont {Shi}, \citenamefont {Goldhaber-Gordon},
  \citenamefont {Zhang},\ and\ \citenamefont {Wang}}]{Chen2019}%
  \BibitemOpen
  \bibfield  {author} {\bibinfo {author} {\bibfnamefont {G.}~\bibnamefont
  {Chen}}, \bibinfo {author} {\bibfnamefont {A.~L.}\ \bibnamefont {Sharpe}},
  \bibinfo {author} {\bibfnamefont {P.}~\bibnamefont {Gallagher}}, \bibinfo
  {author} {\bibfnamefont {I.~T.}\ \bibnamefont {Rosen}}, \bibinfo {author}
  {\bibfnamefont {E.~J.}\ \bibnamefont {Fox}}, \bibinfo {author} {\bibfnamefont
  {L.}~\bibnamefont {Jiang}}, \bibinfo {author} {\bibfnamefont
  {B.}~\bibnamefont {Lyu}}, \bibinfo {author} {\bibfnamefont {H.}~\bibnamefont
  {Li}}, \bibinfo {author} {\bibfnamefont {K.}~\bibnamefont {Watanabe}},
  \bibinfo {author} {\bibfnamefont {T.}~\bibnamefont {Taniguchi}}, \bibinfo
  {author} {\bibfnamefont {J.}~\bibnamefont {Jung}}, \bibinfo {author}
  {\bibfnamefont {Z.}~\bibnamefont {Shi}}, \bibinfo {author} {\bibfnamefont
  {D.}~\bibnamefont {Goldhaber-Gordon}}, \bibinfo {author} {\bibfnamefont
  {Y.}~\bibnamefont {Zhang}}, \ and\ \bibinfo {author} {\bibfnamefont
  {F.}~\bibnamefont {Wang}},\ }\href {\doibase 10.1038/s41586-019-1393-y}
  {\bibfield  {journal} {\bibinfo  {journal} {Nature}\ }\textbf {\bibinfo
  {volume} {572}},\ \bibinfo {pages} {215} (\bibinfo {year}
  {2019})}\BibitemShut {NoStop}%
\bibitem [{\citenamefont {{Lu}}\ \emph {et~al.}(2019)\citenamefont {{Lu}},
  \citenamefont {{Stepanov}}, \citenamefont {{Yang}}, \citenamefont {{Xie}},
  \citenamefont {{Aamir}}, \citenamefont {{Das}}, \citenamefont {{Urgell}},
  \citenamefont {{Watanabe}}, \citenamefont {{Taniguchi}}, \citenamefont
  {{Zhang}}, \citenamefont {{Bachtold}}, \citenamefont {{MacDonald}},\ and\
  \citenamefont {{Efetov}}}]{Lu2019arXiv}%
  \BibitemOpen
  \bibfield  {author} {\bibinfo {author} {\bibfnamefont {X.}~\bibnamefont
  {{Lu}}}, \bibinfo {author} {\bibfnamefont {P.}~\bibnamefont {{Stepanov}}},
  \bibinfo {author} {\bibfnamefont {W.}~\bibnamefont {{Yang}}}, \bibinfo
  {author} {\bibfnamefont {M.}~\bibnamefont {{Xie}}}, \bibinfo {author}
  {\bibfnamefont {M.~A.}\ \bibnamefont {{Aamir}}}, \bibinfo {author}
  {\bibfnamefont {I.}~\bibnamefont {{Das}}}, \bibinfo {author} {\bibfnamefont
  {C.}~\bibnamefont {{Urgell}}}, \bibinfo {author} {\bibfnamefont
  {K.}~\bibnamefont {{Watanabe}}}, \bibinfo {author} {\bibfnamefont
  {T.}~\bibnamefont {{Taniguchi}}}, \bibinfo {author} {\bibfnamefont
  {G.}~\bibnamefont {{Zhang}}}, \bibinfo {author} {\bibfnamefont
  {A.}~\bibnamefont {{Bachtold}}}, \bibinfo {author} {\bibfnamefont {A.~H.}\
  \bibnamefont {{MacDonald}}}, \ and\ \bibinfo {author} {\bibfnamefont {D.~K.}\
  \bibnamefont {{Efetov}}},\ }\href {\doibase 10.1038/s41586-019-1695-0}
  {\bibfield  {journal} {\bibinfo  {journal} {Nature}\ }\textbf {\bibinfo
  {volume} {574}},\ \bibinfo {pages} {653} (\bibinfo {year}
  {2019})} \BibitemShut {NoStop}%
\bibitem [{\citenamefont {{Shen}}\ \emph {et~al.}(2019)\citenamefont {{Shen}},
  \citenamefont {{Li}}, \citenamefont {{Wang}}, \citenamefont {{Zhao}},
  \citenamefont {{Tang}}, \citenamefont {{Liu}}, \citenamefont {{Tian}},
  \citenamefont {{Chu}}, \citenamefont {{Watanabe}}, \citenamefont
  {{Taniguchi}}, \citenamefont {Yang}, \citenamefont {Meng}, \citenamefont
  {Shi},\ and\ \citenamefont {Zhang}}]{Shen2019arXiv190306952S}%
  \BibitemOpen
  \bibfield  {author} {\bibinfo {author} {\bibfnamefont {C.}~\bibnamefont
  {{Shen}}}, \bibinfo {author} {\bibfnamefont {N.}~\bibnamefont {{Li}}},
  \bibinfo {author} {\bibfnamefont {S.}~\bibnamefont {{Wang}}}, \bibinfo
  {author} {\bibfnamefont {Y.}~\bibnamefont {{Zhao}}}, \bibinfo {author}
  {\bibfnamefont {J.}~\bibnamefont {{Tang}}}, \bibinfo {author} {\bibfnamefont
  {J.}~\bibnamefont {{Liu}}}, \bibinfo {author} {\bibfnamefont
  {J.}~\bibnamefont {{Tian}}}, \bibinfo {author} {\bibfnamefont
  {Y.}~\bibnamefont {{Chu}}}, \bibinfo {author} {\bibfnamefont
  {K.}~\bibnamefont {{Watanabe}}}, \bibinfo {author} {\bibfnamefont
  {T.}~\bibnamefont {{Taniguchi}}}, \bibinfo {author} {\bibfnamefont
  {R.}~\bibnamefont {Yang}}, \bibinfo {author} {\bibfnamefont {Z.~Y.}\
  \bibnamefont {Meng}}, \bibinfo {author} {\bibfnamefont {D.}~\bibnamefont
  {Shi}}, \ and\ \bibinfo {author} {\bibfnamefont {G.}~\bibnamefont {Zhang}},\
  }\href@noop {} {\bibfield  {journal} {\bibinfo  {journal} {arXiv e-prints}\
  ,\ \bibinfo {eid} {arXiv:1903.06952}} (\bibinfo {year} {2019})},\ \Eprint
  {http://arxiv.org/abs/1903.06952} {arXiv:1903.06952 [cond-mat.supr-con]}
  \BibitemShut {NoStop}%
\bibitem [{\citenamefont {{Liu}}\ \emph
  {et~al.}(2019{\natexlab{b}})\citenamefont {{Liu}}, \citenamefont {{Hao}},
  \citenamefont {{Khalaf}}, \citenamefont {{Lee}}, \citenamefont {{Watanabe}},
  \citenamefont {{Taniguchi}}, \citenamefont {{Vishwanath}},\ and\
  \citenamefont {{Kim}}}]{Kim2019}%
  \BibitemOpen
  \bibfield  {author} {\bibinfo {author} {\bibfnamefont {X.}~\bibnamefont
  {{Liu}}}, \bibinfo {author} {\bibfnamefont {Z.}~\bibnamefont {{Hao}}},
  \bibinfo {author} {\bibfnamefont {E.}~\bibnamefont {{Khalaf}}}, \bibinfo
  {author} {\bibfnamefont {J.~Y.}\ \bibnamefont {{Lee}}}, \bibinfo {author}
  {\bibfnamefont {K.}~\bibnamefont {{Watanabe}}}, \bibinfo {author}
  {\bibfnamefont {T.}~\bibnamefont {{Taniguchi}}}, \bibinfo {author}
  {\bibfnamefont {A.}~\bibnamefont {{Vishwanath}}}, \ and\ \bibinfo {author}
  {\bibfnamefont {P.}~\bibnamefont {{Kim}}},\ }\href@noop {} {\bibfield
  {journal} {\bibinfo  {journal} {arXiv e-prints}\ ,\ \bibinfo {eid}
  {arXiv:1903.08130}} (\bibinfo {year} {2019}{\natexlab{b}})},\ \Eprint
  {http://arxiv.org/abs/1903.08130} {arXiv:1903.08130 [cond-mat.mes-hall]}
  \BibitemShut {NoStop}%
\bibitem [{\citenamefont {{Cao}}\ \emph {et~al.}(2019)\citenamefont {{Cao}},
  \citenamefont {{Rodan-Legrain}}, \citenamefont {{Rubies-Bigord{\`a}}},
  \citenamefont {{Park}}, \citenamefont {{Watanabe}}, \citenamefont
  {{Taniguchi}},\ and\ \citenamefont
  {{Jarillo-Herrero}}}]{Cao2019arXiv190308596C}%
  \BibitemOpen
  \bibfield  {author} {\bibinfo {author} {\bibfnamefont {Y.}~\bibnamefont
  {{Cao}}}, \bibinfo {author} {\bibfnamefont {D.}~\bibnamefont
  {{Rodan-Legrain}}}, \bibinfo {author} {\bibfnamefont {O.}~\bibnamefont
  {{Rubies-Bigord{\`a}}}}, \bibinfo {author} {\bibfnamefont {J.~M.}\
  \bibnamefont {{Park}}}, \bibinfo {author} {\bibfnamefont {K.}~\bibnamefont
  {{Watanabe}}}, \bibinfo {author} {\bibfnamefont {T.}~\bibnamefont
  {{Taniguchi}}}, \ and\ \bibinfo {author} {\bibfnamefont {P.}~\bibnamefont
  {{Jarillo-Herrero}}},\ }\href@noop {} {\bibfield  {journal} {\bibinfo
  {journal} {arXiv e-prints}\ ,\ \bibinfo {eid} {arXiv:1903.08596}} (\bibinfo
  {year} {2019})},\ \Eprint {http://arxiv.org/abs/1903.08596} {arXiv:1903.08596
  [cond-mat.str-el]} \BibitemShut {NoStop}%
\bibitem [{\citenamefont {{Chen}}\ \emph {et~al.}(2019)\citenamefont {{Chen}},
  \citenamefont {{Sharpe}}, \citenamefont {{Fox}}, \citenamefont {{Zhang}},
  \citenamefont {{Wang}}, \citenamefont {{Jiang}}, \citenamefont {{Lyu}},
  \citenamefont {{Li}}, \citenamefont {{Watanabe}}, \citenamefont
  {{Taniguchi}}, \citenamefont {Shi}, \citenamefont {Senthil}, \citenamefont
  {Goldhaber-Gordon}, \citenamefont {Zhang},\ and\ \citenamefont
  {Wang}}]{Chen2019arXiv190506535C}%
  \BibitemOpen
  \bibfield  {author} {\bibinfo {author} {\bibfnamefont {G.}~\bibnamefont
  {{Chen}}}, \bibinfo {author} {\bibfnamefont {A.~L.}\ \bibnamefont
  {{Sharpe}}}, \bibinfo {author} {\bibfnamefont {E.~J.}\ \bibnamefont {{Fox}}},
  \bibinfo {author} {\bibfnamefont {Y.-H.}\ \bibnamefont {{Zhang}}}, \bibinfo
  {author} {\bibfnamefont {S.}~\bibnamefont {{Wang}}}, \bibinfo {author}
  {\bibfnamefont {L.}~\bibnamefont {{Jiang}}}, \bibinfo {author} {\bibfnamefont
  {B.}~\bibnamefont {{Lyu}}}, \bibinfo {author} {\bibfnamefont
  {H.}~\bibnamefont {{Li}}}, \bibinfo {author} {\bibfnamefont {K.}~\bibnamefont
  {{Watanabe}}}, \bibinfo {author} {\bibfnamefont {T.}~\bibnamefont
  {{Taniguchi}}}, \bibinfo {author} {\bibfnamefont {Z.}~\bibnamefont {Shi}},
  \bibinfo {author} {\bibfnamefont {T.}~\bibnamefont {Senthil}}, \bibinfo
  {author} {\bibfnamefont {D.}~\bibnamefont {Goldhaber-Gordon}}, \bibinfo
  {author} {\bibfnamefont {Y.}~\bibnamefont {Zhang}}, \ and\ \bibinfo {author}
  {\bibfnamefont {F.}~\bibnamefont {Wang}},\ }\href@noop {} {\bibfield
  {journal} {\bibinfo  {journal} {arXiv e-prints}\ ,\ \bibinfo {eid}
  {arXiv:1905.06535}} (\bibinfo {year} {2019})},\ \Eprint
  {http://arxiv.org/abs/1905.06535} {arXiv:1905.06535 [cond-mat.mes-hall]}
  \BibitemShut {NoStop}%
\bibitem [{\citenamefont {Peotta}\ and\ \citenamefont
  {T{\"{o}}rm{\"{a}}}(2015)}]{Peotta2015}%
  \BibitemOpen
  \bibfield  {author} {\bibinfo {author} {\bibfnamefont {S.}~\bibnamefont
  {Peotta}}\ and\ \bibinfo {author} {\bibfnamefont {P.}~\bibnamefont
  {T{\"{o}}rm{\"{a}}}},\ }\href {\doibase 10.1038/ncomms9944} {\bibfield
  {journal} {\bibinfo  {journal} {Nat. Commun.}\ }\textbf {\bibinfo {volume}
  {6}},\ \bibinfo {pages} {8944} (\bibinfo {year} {2015})},\ \Eprint
  {http://arxiv.org/abs/1506.02815} {arXiv:1506.02815} \BibitemShut {NoStop}%
\bibitem [{\citenamefont {Liang}\ \emph {et~al.}(2017)\citenamefont {Liang},
  \citenamefont {Vanhala}, \citenamefont {Peotta}, \citenamefont {Siro},
  \citenamefont {Harju},\ and\ \citenamefont {T{\"{o}}rm{\"{a}}}}]{Liang2017}%
  \BibitemOpen
  \bibfield  {author} {\bibinfo {author} {\bibfnamefont {L.}~\bibnamefont
  {Liang}}, \bibinfo {author} {\bibfnamefont {T.~I.}\ \bibnamefont {Vanhala}},
  \bibinfo {author} {\bibfnamefont {S.}~\bibnamefont {Peotta}}, \bibinfo
  {author} {\bibfnamefont {T.}~\bibnamefont {Siro}}, \bibinfo {author}
  {\bibfnamefont {A.}~\bibnamefont {Harju}}, \ and\ \bibinfo {author}
  {\bibfnamefont {P.}~\bibnamefont {T{\"{o}}rm{\"{a}}}},\ }\href {\doibase
  10.1103/PhysRevB.95.024515} {\bibfield  {journal} {\bibinfo  {journal} {Phys.
  Rev. B}\ }\textbf {\bibinfo {volume} {95}},\ \bibinfo {pages} {024515}
  (\bibinfo {year} {2017})},\ \Eprint {http://arxiv.org/abs/1610.01803}
  {arXiv:1610.01803} \BibitemShut {NoStop}%
\bibitem [{\citenamefont {Hazra}\ \emph {et~al.}(2019)\citenamefont {Hazra},
  \citenamefont {Verma},\ and\ \citenamefont {Randeria}}]{Hazra2018}%
  \BibitemOpen
  \bibfield  {author} {\bibinfo {author} {\bibfnamefont {T.}~\bibnamefont
  {Hazra}}, \bibinfo {author} {\bibfnamefont {N.}~\bibnamefont {Verma}}, \ and\
  \bibinfo {author} {\bibfnamefont {M.}~\bibnamefont {Randeria}},\ }\href
  {\doibase 10.1103/PhysRevX.9.031049} {\bibfield  {journal} {\bibinfo
  {journal} {Phys. Rev. X}\ }\textbf {\bibinfo {volume} {9}},\ \bibinfo {pages}
  {031049} (\bibinfo {year} {2019})}\BibitemShut {NoStop}%
\bibitem [{\citenamefont {{Xie}}\ \emph {et~al.}(2019)\citenamefont {{Xie}},
  \citenamefont {{Song}}, \citenamefont {{Lian}},\ and\ \citenamefont
  {{Bernevig}}}]{Fang2019}%
  \BibitemOpen
  \bibfield  {author} {\bibinfo {author} {\bibfnamefont {F.}~\bibnamefont
  {{Xie}}}, \bibinfo {author} {\bibfnamefont {Z.}~\bibnamefont {{Song}}},
  \bibinfo {author} {\bibfnamefont {B.}~\bibnamefont {{Lian}}}, \ and\ \bibinfo
  {author} {\bibfnamefont {B.~A.}\ \bibnamefont {{Bernevig}}},\ }\href@noop {}
  {\bibfield  {journal} {\bibinfo  {journal} {arXiv e-prints}\ ,\ \bibinfo
  {eid} {arXiv:1906.02213}} (\bibinfo {year} {2019})},\ \Eprint
  {http://arxiv.org/abs/1906.02213} {arXiv:1906.02213 [cond-mat.supr-con]}
  \BibitemShut {NoStop}%
\bibitem [{\citenamefont {Jung}\ \emph {et~al.}(2014)\citenamefont {Jung},
  \citenamefont {Raoux}, \citenamefont {Qiao},\ and\ \citenamefont
  {MacDonald}}]{Jung2014}%
  \BibitemOpen
  \bibfield  {author} {\bibinfo {author} {\bibfnamefont {J.}~\bibnamefont
  {Jung}}, \bibinfo {author} {\bibfnamefont {A.}~\bibnamefont {Raoux}},
  \bibinfo {author} {\bibfnamefont {Z.}~\bibnamefont {Qiao}}, \ and\ \bibinfo
  {author} {\bibfnamefont {A.~H.}\ \bibnamefont {MacDonald}},\ }\href {\doibase
  10.1103/PhysRevB.89.205414} {\bibfield  {journal} {\bibinfo  {journal} {Phys.
  Rev. B}\ }\textbf {\bibinfo {volume} {89}},\ \bibinfo {pages} {205414}
  (\bibinfo {year} {2014})}\BibitemShut {NoStop}%
\bibitem [{SM()}]{SM}%
  \BibitemOpen
  \href@noop {} {}\bibinfo {note} {See supplemental material.}\BibitemShut
  {Stop}%
\bibitem [{\citenamefont {Scalapino}\ \emph {et~al.}(1992)\citenamefont
  {Scalapino}, \citenamefont {White},\ and\ \citenamefont
  {Zhang}}]{Scalapino1992}%
  \BibitemOpen
  \bibfield  {author} {\bibinfo {author} {\bibfnamefont {D.~J.}\ \bibnamefont
  {Scalapino}}, \bibinfo {author} {\bibfnamefont {S.~R.}\ \bibnamefont
  {White}}, \ and\ \bibinfo {author} {\bibfnamefont {S.~C.}\ \bibnamefont
  {Zhang}},\ }\href {\doibase 10.1103/PhysRevLett.68.2830} {\bibfield
  {journal} {\bibinfo  {journal} {Phys. Rev. Lett.}\ }\textbf {\bibinfo
  {volume} {68}},\ \bibinfo {pages} {2830} (\bibinfo {year}
  {1992})}\BibitemShut {NoStop}%
\bibitem [{\citenamefont {Scalapino}\ \emph {et~al.}(1993)\citenamefont
  {Scalapino}, \citenamefont {White},\ and\ \citenamefont
  {Zhang}}]{Scalapino1993}%
  \BibitemOpen
  \bibfield  {author} {\bibinfo {author} {\bibfnamefont {D.~J.}\ \bibnamefont
  {Scalapino}}, \bibinfo {author} {\bibfnamefont {S.~R.}\ \bibnamefont
  {White}}, \ and\ \bibinfo {author} {\bibfnamefont {S.}~\bibnamefont
  {Zhang}},\ }\href {\doibase 10.1103/PhysRevB.47.7995} {\bibfield  {journal}
  {\bibinfo  {journal} {Phys. Rev. B}\ }\textbf {\bibinfo {volume} {47}},\
  \bibinfo {pages} {7995} (\bibinfo {year} {1993})}\BibitemShut {NoStop}%
\bibitem [{\citenamefont {Kerelsky}\ \emph {et~al.}(2019)\citenamefont
  {Kerelsky}, \citenamefont {McGilly}, \citenamefont {Kennes}, \citenamefont
  {Xian}, \citenamefont {Yankowitz}, \citenamefont {Chen}, \citenamefont
  {Watanabe}, \citenamefont {Taniguchi}, \citenamefont {Hone}, \citenamefont
  {Dean}, \citenamefont {Rubio},\ and\ \citenamefont
  {Pasupathy}}]{Kerelsky2019}%
  \BibitemOpen
  \bibfield  {author} {\bibinfo {author} {\bibfnamefont {A.}~\bibnamefont
  {Kerelsky}}, \bibinfo {author} {\bibfnamefont {L.~J.}\ \bibnamefont
  {McGilly}}, \bibinfo {author} {\bibfnamefont {D.~M.}\ \bibnamefont {Kennes}},
  \bibinfo {author} {\bibfnamefont {L.}~\bibnamefont {Xian}}, \bibinfo {author}
  {\bibfnamefont {M.}~\bibnamefont {Yankowitz}}, \bibinfo {author}
  {\bibfnamefont {S.}~\bibnamefont {Chen}}, \bibinfo {author} {\bibfnamefont
  {K.}~\bibnamefont {Watanabe}}, \bibinfo {author} {\bibfnamefont
  {T.}~\bibnamefont {Taniguchi}}, \bibinfo {author} {\bibfnamefont
  {J.}~\bibnamefont {Hone}}, \bibinfo {author} {\bibfnamefont {C.}~\bibnamefont
  {Dean}}, \bibinfo {author} {\bibfnamefont {A.}~\bibnamefont {Rubio}}, \ and\
  \bibinfo {author} {\bibfnamefont {A.~N.}\ \bibnamefont {Pasupathy}},\ }\href
  {\doibase 10.1038/s41586-019-1431-9} {\bibfield  {journal} {\bibinfo
  {journal} {Nature}\ }\textbf {\bibinfo {volume} {572}},\ \bibinfo {pages}
  {95} (\bibinfo {year} {2019})}\BibitemShut {NoStop}%
\bibitem [{\citenamefont {Rhim}\ and\ \citenamefont {Yang}(2019)}]{Rhim2019}%
  \BibitemOpen
  \bibfield  {author} {\bibinfo {author} {\bibfnamefont {J.-W.}\ \bibnamefont
  {Rhim}}\ and\ \bibinfo {author} {\bibfnamefont {B.-J.}\ \bibnamefont
  {Yang}},\ }\href {\doibase 10.1103/PhysRevB.99.045107} {\bibfield  {journal}
  {\bibinfo  {journal} {Phys. Rev. B}\ }\textbf {\bibinfo {volume} {99}},\
  \bibinfo {pages} {045107} (\bibinfo {year} {2019})}\BibitemShut {NoStop}%
\bibitem [{\citenamefont {{Berezinski{\v{i}}}}(1971)}]{Berezinski1971}%
  \BibitemOpen
  \bibfield  {author} {\bibinfo {author} {\bibfnamefont {V.~L.}\ \bibnamefont
  {{Berezinski{\v{i}}}}},\ }\href
  {http://www.jetp.ac.ru/cgi-bin/e/index/e/32/3/p493?a=list} {\bibfield
  {journal} {\bibinfo  {journal} {Zh. Eksp. Teor. Fiz.}\ }\textbf {\bibinfo {volume} {59}},\ \bibinfo {pages}
  {907} (\bibinfo {year} {1970}) [Sov.
J. Exp. Theor. Phys. \textbf{32}, 493 (1971)]}\BibitemShut {NoStop}%
\bibitem [{\citenamefont {Kosterlitz}\ and\ \citenamefont
  {Thouless}(1973)}]{Kosterlitz1973}%
  \BibitemOpen
  \bibfield  {author} {\bibinfo {author} {\bibfnamefont {J.~M.}\ \bibnamefont
  {Kosterlitz}}\ and\ \bibinfo {author} {\bibfnamefont {D.~J.}\ \bibnamefont
  {Thouless}},\ }\href {\doibase 10.1088/0022-3719/6/7/010} {\bibfield
  {journal} {\bibinfo  {journal} {Journal of Physics C: Solid State Physics}\
  }\textbf {\bibinfo {volume} {6}},\ \bibinfo {pages} {1181} (\bibinfo {year}
  {1973})}\BibitemShut {NoStop}%
\bibitem [{\citenamefont {Hebard}\ and\ \citenamefont
  {Fiory}(1980)}]{Hebard1980}%
  \BibitemOpen
  \bibfield  {author} {\bibinfo {author} {\bibfnamefont {A.~F.}\ \bibnamefont
  {Hebard}}\ and\ \bibinfo {author} {\bibfnamefont {A.~T.}\ \bibnamefont
  {Fiory}},\ }\href {\doibase 10.1103/PhysRevLett.44.291} {\bibfield  {journal}
  {\bibinfo  {journal} {Phys. Rev. Lett.}\ }\textbf {\bibinfo {volume} {44}},\
  \bibinfo {pages} {291} (\bibinfo {year} {1980})}\BibitemShut {NoStop}%
\bibitem [{\citenamefont {Turneaure}\ \emph {et~al.}(2000)\citenamefont
  {Turneaure}, \citenamefont {Lemberger},\ and\ \citenamefont
  {Graybeal}}]{Turneaure2000}%
  \BibitemOpen
  \bibfield  {author} {\bibinfo {author} {\bibfnamefont {S.~J.}\ \bibnamefont
  {Turneaure}}, \bibinfo {author} {\bibfnamefont {T.~R.}\ \bibnamefont
  {Lemberger}}, \ and\ \bibinfo {author} {\bibfnamefont {J.~M.}\ \bibnamefont
  {Graybeal}},\ }\href {\doibase 10.1103/PhysRevLett.84.987} {\bibfield
  {journal} {\bibinfo  {journal} {Phys. Rev. Lett.}\ }\textbf {\bibinfo
  {volume} {84}},\ \bibinfo {pages} {987} (\bibinfo {year} {2000})}\BibitemShut
  {NoStop}%
\bibitem [{\citenamefont {Bert}\ \emph {et~al.}(2011)\citenamefont {Bert},
  \citenamefont {Kalisky}, \citenamefont {Bell}, \citenamefont {Kim},
  \citenamefont {Hikita}, \citenamefont {Hwang},\ and\ \citenamefont
  {Moler}}]{Bert2011}%
  \BibitemOpen
  \bibfield  {author} {\bibinfo {author} {\bibfnamefont {J.~A.}\ \bibnamefont
  {Bert}}, \bibinfo {author} {\bibfnamefont {B.}~\bibnamefont {Kalisky}},
  \bibinfo {author} {\bibfnamefont {C.}~\bibnamefont {Bell}}, \bibinfo {author}
  {\bibfnamefont {M.}~\bibnamefont {Kim}}, \bibinfo {author} {\bibfnamefont
  {Y.}~\bibnamefont {Hikita}}, \bibinfo {author} {\bibfnamefont {H.~Y.}\
  \bibnamefont {Hwang}}, \ and\ \bibinfo {author} {\bibfnamefont {K.~A.}\
  \bibnamefont {Moler}},\ }\href {\doibase 10.1038/nphys2079} {\bibfield
  {journal} {\bibinfo  {journal} {Nat. Phys.}\ }\textbf {\bibinfo {volume}
  {7}},\ \bibinfo {pages} {767} (\bibinfo {year} {2011})}\BibitemShut {NoStop}%
\bibitem [{\citenamefont {Bert}\ \emph {et~al.}(2012)\citenamefont {Bert},
  \citenamefont {Nowack}, \citenamefont {Kalisky}, \citenamefont {Noad},
  \citenamefont {Kirtley}, \citenamefont {Bell}, \citenamefont {Sato},
  \citenamefont {Hosoda}, \citenamefont {Hikita}, \citenamefont {Hwang},\ and\
  \citenamefont {Moler}}]{Bert2012}%
  \BibitemOpen
  \bibfield  {author} {\bibinfo {author} {\bibfnamefont {J.~A.}\ \bibnamefont
  {Bert}}, \bibinfo {author} {\bibfnamefont {K.~C.}\ \bibnamefont {Nowack}},
  \bibinfo {author} {\bibfnamefont {B.}~\bibnamefont {Kalisky}}, \bibinfo
  {author} {\bibfnamefont {H.}~\bibnamefont {Noad}}, \bibinfo {author}
  {\bibfnamefont {J.~R.}\ \bibnamefont {Kirtley}}, \bibinfo {author}
  {\bibfnamefont {C.}~\bibnamefont {Bell}}, \bibinfo {author} {\bibfnamefont
  {H.~K.}\ \bibnamefont {Sato}}, \bibinfo {author} {\bibfnamefont
  {M.}~\bibnamefont {Hosoda}}, \bibinfo {author} {\bibfnamefont
  {Y.}~\bibnamefont {Hikita}}, \bibinfo {author} {\bibfnamefont {H.~Y.}\
  \bibnamefont {Hwang}}, \ and\ \bibinfo {author} {\bibfnamefont {K.~A.}\
  \bibnamefont {Moler}},\ }\href {\doibase 10.1103/PhysRevB.86.060503}
  {\bibfield  {journal} {\bibinfo  {journal} {Phys. Rev. B}\ }\textbf {\bibinfo
  {volume} {86}},\ \bibinfo {pages} {060503(R)} (\bibinfo {year}
  {2012})}\BibitemShut {NoStop}%
\bibitem [{\citenamefont {Kapon}\ \emph {et~al.}(2019)\citenamefont {Kapon},
  \citenamefont {Salman}, \citenamefont {Mangel}, \citenamefont {Prokscha},
  \citenamefont {Gavish},\ and\ \citenamefont {Keren}}]{Kapon2019}%
  \BibitemOpen
  \bibfield  {author} {\bibinfo {author} {\bibfnamefont {I.}~\bibnamefont
  {Kapon}}, \bibinfo {author} {\bibfnamefont {Z.}~\bibnamefont {Salman}},
  \bibinfo {author} {\bibfnamefont {I.}~\bibnamefont {Mangel}}, \bibinfo
  {author} {\bibfnamefont {T.}~\bibnamefont {Prokscha}}, \bibinfo {author}
  {\bibfnamefont {N.}~\bibnamefont {Gavish}}, \ and\ \bibinfo {author}
  {\bibfnamefont {A.}~\bibnamefont {Keren}},\ }\href {\doibase
  10.1038/s41467-019-10480-x} {\bibfield  {journal} {\bibinfo  {journal} {Nat.
  Commun.}\ }\textbf {\bibinfo {volume} {10}},\ \bibinfo {pages} {2463}
  (\bibinfo {year} {2019})}\BibitemShut {NoStop}%
\bibitem [{\citenamefont {Julku}\ \emph {et~al.}(2019)\citenamefont {Julku},
  \citenamefont {Peltonen}, \citenamefont {Liang}, \citenamefont
  {Heikkil{\"{a}}},\ and\ \citenamefont {T{\"{o}}rm{\"{a}}}}]{Julku2019}%
  \BibitemOpen
  \bibfield  {author} {\bibinfo {author} {\bibfnamefont {A.}~\bibnamefont
  {Julku}}, \bibinfo {author} {\bibfnamefont {T.~J.}\ \bibnamefont {Peltonen}},
  \bibinfo {author} {\bibfnamefont {L.}~\bibnamefont {Liang}}, \bibinfo
  {author} {\bibfnamefont {T.~T.}\ \bibnamefont {Heikkil{\"{a}}}}, \ and\
  \bibinfo {author} {\bibfnamefont {P.}~\bibnamefont {T{\"{o}}rm{\"{a}}}},\
  }\href@noop {} {\bibfield  {journal} {\bibinfo  {journal} {arXiv e-prints}\
  ,\ \bibinfo {pages} {arXiv:1906.06313v2}} (\bibinfo {year} {2019})},\ \Eprint
  {http://arxiv.org/abs/arXiv:1906.06313v2} {arXiv:arXiv:1906.06313v2}
  \BibitemShut {NoStop}%
\end{thebibliography}

\begin{thebibliography}{4}%
\makeatletter
\providecommand \@ifxundefined [1]{%
 \@ifx{#1\undefined}
}%
\providecommand \@ifnum [1]{%
 \ifnum #1\expandafter \@firstoftwo
 \else \expandafter \@secondoftwo
 \fi
}%
\providecommand \@ifx [1]{%
 \ifx #1\expandafter \@firstoftwo
 \else \expandafter \@secondoftwo
 \fi
}%
\providecommand \natexlab [1]{#1}%
\providecommand \enquote  [1]{``#1''}%
\providecommand \bibnamefont  [1]{#1}%
\providecommand \bibfnamefont [1]{#1}%
\providecommand \citenamefont [1]{#1}%
\providecommand \href@noop [0]{\@secondoftwo}%
\providecommand \href [0]{\begingroup \@sanitize@url \@href}%
\providecommand \@href[1]{\@@startlink{#1}\@@href}%
\providecommand \@@href[1]{\endgroup#1\@@endlink}%
\providecommand \@sanitize@url [0]{\catcode `\\12\catcode `\$12\catcode
  `\&12\catcode `\#12\catcode `\^12\catcode `\_12\catcode `\%12\relax}%
\providecommand \@@startlink[1]{}%
\providecommand \@@endlink[0]{}%
\providecommand \url  [0]{\begingroup\@sanitize@url \@url }%
\providecommand \@url [1]{\endgroup\@href {#1}{\urlprefix }}%
\providecommand \urlprefix  [0]{URL }%
\providecommand \Eprint [0]{\href }%
\providecommand \doibase [0]{http://dx.doi.org/}%
\providecommand \selectlanguage [0]{\@gobble}%
\providecommand \bibinfo  [0]{\@secondoftwo}%
\providecommand \bibfield  [0]{\@secondoftwo}%
\providecommand \translation [1]{[#1]}%
\providecommand \BibitemOpen [0]{}%
\providecommand \bibitemStop [0]{}%
\providecommand \bibitemNoStop [0]{.\EOS\space}%
\providecommand \EOS [0]{\spacefactor3000\relax}%
\providecommand \BibitemShut  [1]{\csname bibitem#1\endcsname}%
\let\auto@bib@innerbib\@empty
\bibitem [{\citenamefont {Wu}\ \emph {et~al.}(2018)\citenamefont {Wu},
  \citenamefont {MacDonald},\ and\ \citenamefont {Martin}}]{Wu2018S}%
  \BibitemOpen
  \bibfield  {author} {\bibinfo {author} {\bibfnamefont {F.}~\bibnamefont
  {Wu}}, \bibinfo {author} {\bibfnamefont {A.~H.}\ \bibnamefont {MacDonald}}, \
  and\ \bibinfo {author} {\bibfnamefont {I.}~\bibnamefont {Martin}},\ }\href
  {\doibase 10.1103/PhysRevLett.121.257001} {\bibfield  {journal} {\bibinfo
  {journal} {Phys. Rev. Lett.}\ }\textbf {\bibinfo {volume} {121}},\ \bibinfo
  {pages} {257001} (\bibinfo {year} {2018})},\ \Eprint
  {http://arxiv.org/abs/1805.08735} {arXiv:1805.08735} \BibitemShut {NoStop}%
\bibitem [{\citenamefont {Liang}\ \emph {et~al.}(2017)\citenamefont {Liang},
  \citenamefont {Vanhala}, \citenamefont {Peotta}, \citenamefont {Siro},
  \citenamefont {Harju},\ and\ \citenamefont {T{\"{o}}rm{\"{a}}}}]{Liang2017S}%
  \BibitemOpen
  \bibfield  {author} {\bibinfo {author} {\bibfnamefont {L.}~\bibnamefont
  {Liang}}, \bibinfo {author} {\bibfnamefont {T.~I.}\ \bibnamefont {Vanhala}},
  \bibinfo {author} {\bibfnamefont {S.}~\bibnamefont {Peotta}}, \bibinfo
  {author} {\bibfnamefont {T.}~\bibnamefont {Siro}}, \bibinfo {author}
  {\bibfnamefont {A.}~\bibnamefont {Harju}}, \ and\ \bibinfo {author}
  {\bibfnamefont {P.}~\bibnamefont {T{\"{o}}rm{\"{a}}}},\ }\href {\doibase
  10.1103/PhysRevB.95.024515} {\bibfield  {journal} {\bibinfo  {journal} {Phys.
  Rev. B}\ }\textbf {\bibinfo {volume} {95}},\ \bibinfo {pages} {024515}
  (\bibinfo {year} {2017})},\ \Eprint {http://arxiv.org/abs/1610.01803}
  {arXiv:1610.01803} \BibitemShut {NoStop}%
\bibitem [{\citenamefont {Bistritzer}\ and\ \citenamefont
  {MacDonald}(2011)}]{Bistritzer2011aS}%
  \BibitemOpen
  \bibfield  {author} {\bibinfo {author} {\bibfnamefont {R.}~\bibnamefont
  {Bistritzer}}\ and\ \bibinfo {author} {\bibfnamefont {A.~H.}\ \bibnamefont
  {MacDonald}},\ }\href {\doibase 10.1073/pnas.1108174108} {\bibfield
  {journal} {\bibinfo  {journal} {Proc. Natl. Acad. Sci. U.S.A.}\ }\textbf
  {\bibinfo {volume} {108}},\ \bibinfo {pages} {12233} (\bibinfo {year}
  {2011})}\BibitemShut {NoStop}%
\bibitem [{\citenamefont {Po}\ \emph {et~al.}(2019)\citenamefont {Po},
  \citenamefont {Zou}, \citenamefont {Senthil},\ and\ \citenamefont
  {Vishwanath}}]{Po2018aS}%
  \BibitemOpen
  \bibfield  {author} {\bibinfo {author} {\bibfnamefont {H.~C.}\ \bibnamefont
  {Po}}, \bibinfo {author} {\bibfnamefont {L.}~\bibnamefont {Zou}}, \bibinfo
  {author} {\bibfnamefont {T.}~\bibnamefont {Senthil}}, \ and\ \bibinfo
  {author} {\bibfnamefont {A.}~\bibnamefont {Vishwanath}},\ }\href {\doibase
  10.1103/PhysRevB.99.195455} {\bibfield  {journal} {\bibinfo  {journal} {Phys.
  Rev. B}\ }\textbf {\bibinfo {volume} {99}},\ \bibinfo {pages} {195455}
  (\bibinfo {year} {2019})}\BibitemShut {NoStop}%
\end{thebibliography}

\end{document}